\documentclass[11pt]{article}
\usepackage[utf8]{inputenc}

\usepackage{graphicx} 
\usepackage{amsmath}
\usepackage{psfrag, epsf}
\usepackage{enumerate}  
\usepackage{natbib}
\usepackage{url} 
\usepackage{multirow}
\usepackage{tabularx}
\usepackage{comment}
\usepackage{subfigure}

\usepackage{float}

\usepackage{algorithm,algpseudocode}

\algnewcommand\algorithmicinput{\textbf{Input:}}
\algnewcommand\algorithmicoutput{\textbf{Output:}}
\algnewcommand\Input{\item[\algorithmicinput]}%
\algnewcommand\Output{\item[\algorithmicoutput]}%

\usepackage[margin=1in]{geometry}

\usepackage{amssymb, amsfonts, amsthm, bm}
\usepackage{mathtools, mathrsfs,amsmath}
\usepackage{relsize} 
\usepackage{hyperref}
\usepackage{enumitem}
\usepackage{enumerate}
\usepackage{verbatim}
\usepackage{wrapfig}
\usepackage{booktabs}
\usepackage{comment}

\newtheorem{lemma}{Lemma}

\newtheorem{example}{Example}

\DeclareMathOperator{\E}{\mathbb{E}}

\DeclareMathOperator*{\argmax}{argmax} 

\DeclareMathOperator{\argmin}{argmin}

\usepackage{color}

\begin{document}

\title{
Unsupervised cell    segmentation  
by fast Gaussian Processes}

\author{Laura Baracaldo$^ \dag $, Blythe King$^ \dag $, Haoran Yan$^ \dag $, Yizi Lin$^ \dag $, Nina Miolane$^\ddag  $, Mengyang Gu$^ \dag $\footnote{The first three authors contribute equally. Correspondence should be addressed to Mengyang Gu (\href{mailto:mengyang@pstat.ucsb.edu}{mengyang@pstat.ucsb.edu}).} \\
\\
           \small $^ \dag  $ Department of Statistics and Applied Probability, University of California, Santa Barbara, CA \\
                      \small $^\ddag  $ Department of Electrical and Computer Engineering, University of California, Santa Barbara, CA 
}
\date{}

\maketitle



\begin{abstract}
Cell boundary information is crucial for analyzing  
cell behaviors from time-lapse microscopy videos. Existing supervised cell segmentation tools, such as {\sf ImageJ}, require tuning various parameters and rely on restrictive assumptions about the  shape of the objects. While recent supervised segmentation tools based on convolutional neural networks enhance accuracy, they depend on high-quality labeled images, making them  unsuitable for segmenting new types of objects not in the database. 
We developed a novel unsupervised cell segmentation algorithm based on fast Gaussian processes for noisy microscopy images without the need for parameter tuning or restrictive assumptions about the shape of the object.  We derived robust thresholding criteria adaptive for heterogeneous images containing distinct brightness at different parts  to  separate  objects from the background, and employed watershed segmentation to distinguish touching cell objects. Both simulated studies and real-data analysis of large microscopy images 
demonstrate the scalability and accuracy of our approach compared with the alternatives.
\end{abstract}


%

\section{Introduction}
\label{sec:intro}
The spatial organization of cells, such as orientation, density, and shape, are fundamental to characterizing physical mechanisms and biological functions \cite{eisenbarth2019dendritic, prasad2019cell}, as the arrangement of cells influences their functions, including proliferation and differentiation \cite{liu2007cellular}. For example, high cell density can signal cells to die to maintain tissue homeostasis \cite{eisenhoffer2012crowding}, and cell shape has been linked to DNA synthesis and cell growth \cite{folkman1978role}. 
Advances in microscopy techniques enable detailed video recordings 
of 
cellular motions. Image segmentation approaches have been developed for  microscopy images \cite{meijering2012cell}, where the segmentation results, 
including cell locomotion, alignment, and density, can be used for modeling cell behaviors \cite{selmeczi2005cell,gu2023data,fang2024inverse,khang2025automated}. Therefore, precise cell image segmentation is essential for quantifying cell location, alignment, and morphology, as these statistics enhance our understanding of how cellular spatial organization influences human health outcomes \cite{tsukui2020collagen,barbazan2023cancer}.

Cellular segmentation tools can be broadly classified as 
unsupervised and supervised methods,  
based on whether labeled data is needed for training the models. One of the most popular tools of unsupervised cellular segmentation is the {\sf ImageJ} \cite{abramoff2004image,schneider2012nih,schindelin2012fiji}, 
which enables segmenting both cell nuclei and whole cells, 
and linking cells across frames of a microscopy video via plug-ins such as {\sf TrackMate} \cite{tinevez2017trackmate}. Prior to using {\sf ImageJ} for cell detection and segmentation, a series of user-driven image pre-processing steps must be completed in the program. For instance, the image 
is typically 
converted into a binary image first, based on the threshold chosen either manually or automatically via several global threshold setting methods \cite{nichele2020quantitative,liu2009otsu}. 
Additional image processing steps, such as 
despeckling for noise reduction 
\cite{ferreira2011imagej}, typically need to be run before segmentation. To detect cell objects within an image, the algorithm scans the corresponding binary image until the edge of an object 
is reached. Then, the algorithm traces around the edge of the object until the starting point is reached. This process is repeated for  pixels of each object, or the foreground pixels, in the binary image. The centroid of each cell object can be identified by fitting an ellipse 
for each object using the second-order central moments \cite{schneider2012nih}, which assumes the cell boundary is elliptical.  The parameters for the best-fit ellipse, including the major and minor axes and centroids, and matrices for the object masks for each object boundary, are recorded  in {\sf ImageJ}
\cite{ferreira2011imagej}. Additional functionality, such as extension of elliptical outlines, 
 can be added to the user interface by writing macro code to streamline image analysis or adding third-party plug-ins \cite{schneider2012nih}. However, achieving reasonable segmentation results is often labor-intensive and requires human interaction at each step. Furthermore, the  assumption of the shape of the cells in {\sf ImageJ} is restrictive. Though the ellipse can approximate nuclei of certain cells,  it cannot approximate the shape of cytoplasm or the whole cell of frequently used cells in experiments, such as fibroblasts or epithelial cells.

Supervised cellular segmentation methods, are typically built upon a convolutional neural network structure, such as the U-net  \cite{ronneberger2015u}. A popular tool is $\sf{Cellpose}$, which builds U-nets { for segmenting images that contain one type of cells} that are trained on a dataset of over 70,000 segmented objects, including annotated images that contain numerous types of cells   \cite{stringer2021cellpose}. The model first compresses the image and then reconstructs the image to its original dimensions  through a U-net to capture both local and global features. It further generates a topological map from diffusion simulation to product gradient maps for the model to {learn} the horizontal and vertical gradients of the image. 
The final layer of {\sf{Cellpose}} produces three feature maps, including the horizontal and vertical directional gradients, and the sigmoid transformation of the probability map of the output image from the U-net, and the parameters of the  {\sf{Cellpose}} are then estimated by fitting a  loss function of the gradient vector and binary label of each pixel of the image. 
As the {\sf{Cellpose}} segmentation requires massive training data, it may not work well for detecting novel cell types, where high-labeled image data that may not be readily available. Furthermore, training {\sf{Cellpose}}  and using it for segmenting large images with several million pixels can be computationally expensive. 

In response to the challenges the current methods face, we present a novel unsupervised cell image segmentation method { for segmenting images containing one type of cells}, based on separable Gaussian processes, automated thresholding, and watershed operations for image segmentation. The proposed method has several advantages. First, the separable covariance structure on image data enables us to decompose the large covariance into a Kronecker product of two small covariances, which substantially reduces the computational cost for computing the matrix inversion and the logarithm of the determinant of the covariance matrix. 
Second, 
unlike conventional unsupervised image segmentation methods such as {\sf ImageJ}, which require manual tuning of multiple hyperparameters, including Gaussian blur radius, thresholding method, and minimum object size, our proposed segmentation pipeline is free of manual parameter selection. 
We developed fast algorithms for maximum likelihood estimation of the parameters in GPs and the predictive distribution. The  predictive mean surface naturally facilitates foreground-background separation by applying a data-driven threshold with the optimal threshold automatically estimated based on the 
absolute second difference 
in foreground pixel counts between the thresholds. 
{Third}, the cell masks are segmented using the watershed algorithm, which simulates water flowing through a topographical map of the image to separate cell objects. 
Fourth, our approach is entirely unsupervised, negating any need for annotated training data, in contrast to the supervised methods, such as {\sf{Cellpose}}. Thus, the proposed method is particularly suitable for segmenting new cell types or other biological structures. 
 We studied our methodology by simulated data for noise filtering. We also compared our approaches with the alternatives using real microscopy images for segmenting both the cell nucleus and whole cell in optically dense images,  
 a challenging scenario even when labeled data are available. These studies demonstrate that our approach substantially outperforms the alternatives.

The rest of the article is organized as follows. We discuss GP models of images and a fast algorithm for parameter estimation and predictive distribution in Section \ref{sec:GP}. 
In Section \ref{sec:IS}, we introduce our workflow for cellular image segmentation from Gaussian processes, including image smoothing, binary thresholding, and segmentation. In Section \ref{sec:sim_study}, 
we use simulated data and microscopy images of cells to compare our methods and other alternatives for denoising, including principal component analysis and dynamic mode decomposition. 
We also demonstrate the computational advantage of the fast algorithm against direct computation. 
Finally, we generate cell masks for both the cell nucleus and cytoplasm, or the whole cell, by our methods and {\sf ImageJ} segmentation method to compare their accuracy with the manual labels in Section \ref{sec:rda}. 
Our findings not only present a novel cell segmentation technique but also highlight the applicability of fast Gaussian processes in image processing tasks. 
The data and code of this article are made publicly available: \url{https://github.com/UncertaintyQuantification/cell_segmentation}.

\section{Gaussian process models of images}
\label{sec:GP}
\subsection{The likelihood function and predictive distribution}
\label{subsec:lik_pred_GP}
Let us consider a Gaussian process (GP) model for a two-dimensional (2D) microscopy image, where the  intensity at the pixel location $\mathbf x$ is modeled as $y(\mathbf x)=f(\mathbf x)+\epsilon$, with $f(\cdot)$ being a latent Gaussian process (GP) and  $\epsilon \sim \mathcal N(0, \sigma^2_0 )$ being a Gaussian noise with variance $\sigma^2_0$ \cite{rasmussen2006gaussian}. Consider an $n_1\times n_2$ image $\mathbf Y$, where the $(i,j)$th term is $y(\mathbf x_{i,j})$ for the pixel location $\mathbf x_{i,j}=(x_{i,1},x_{j,2})$. The data matrix can be written as:    
\begin{equation}
\mathbf Y=\mathbf F+\mathbf E, 
\label{equ:model_matrix}
\end{equation}
where $\mathbf F$ follows a matrix normal distribution $\mathbf F \sim \mbox{Matrix-Normal} \left(\mu \mathbf 1_{n_1\times n_2}, \, \sigma^2 \mathbf R_1, \mathbf R_2\right)$ with $\mu$ being a mean parameter, $\mathbf 1_{n_1\times n_2}$ being a $n_1\times n_2$ matrix of ones, $\sigma^2$ being the variance parameter, $\mathbf R_1$ and $\mathbf R_2$ being correlation matrices for two inputs, respectively, and the {{vectorized noise matrix follows
$\mbox{Vec}(\mathbf E) \sim \mathcal{MN}(0,\sigma^2_0 \mathbf I_{N \times N})$, with $N=n_1n_2$.}}  Additional trend structure can be modeled in the mean function, and the covariance can be generalized to be semi-separable in the model \cite{gu2022gaussian}. 

Vectorize the observations matrix $\mathbf y_v=\mbox{Vec}(\mathbf Y)$. 
After marginalizing out the latent signal matrix $\mathbf F$, the observation vector of the image follows a multivariate normal distribution,  
\begin{align}
{
( \mathbf y_v \mid \sigma^2 , \eta, \mathbf R_1, \mathbf R_2) \sim 
\mathcal{MN}(\mathbf \mu \mathbf 1_N,\sigma^2 \mathbf{\tilde R}),
}
\label{equ:marginal_model}
\end{align}
where {{$ \mathbf{\tilde R}=\mathbf R +\eta \mathbf I_{N}$, $\mathbf{R}=\mathbf R_2 \otimes \mathbf R_1 $}}, $\otimes$ is a Kronecker product and $\eta=\sigma^2_0/\sigma^2$ is a nugget parameter. Here the correlation matrices are parameterized kernel functions $ R_{1}(i,i')= K_1(x_{i,1}, x_{i',1})$ and $R_{2}(j,j')=K_2(x_{j,2}, x_{j',2})$. Frequently used kernel functions include the power exponential kernel and Mat{\'e}rn kernel \cite{rasmussen2006gaussian}.  Denote   $K_l(x_{i,l}, x_{i',l})=K_l(d_l)$ with $d_l=|x_{i,l}-x_{i',l}|$. For instance, the Mat{\'e}rn kernel with roughness parameter $5/2$ follows \cite{handcock1993bayesian}:
 \begin{equation}
 K_l(d_l)=\left(1+\frac{\sqrt{5}  d_l}{\gamma_l}+ \frac{5  d^2_l}{3 \gamma^2_l}\right) \exp \left(-\frac{\sqrt{5}  d_l}{\gamma_l}\right),
 \label{equ:matern_5_2}
 \end{equation}
where $\gamma_l$ is a range parameter to be estimated for $l=1,2$. A GP with the kernel function in Equation (\ref{equ:matern_5_2}) is twice mean-squared differentiable \cite{rasmussen2006gaussian}, and it is used as a default choice of some software packages of GP surrogate models \cite{roustant2012dicekriging,gu2018robustgasp}. 
We use the kernel function in 
 (\ref{equ:matern_5_2}) for demonstration purposes.

After specifying the model, we compute the likelihood function and predictive distribution for parameter estimation and prediction, respectively. The parameters of the model in 
 (\ref{equ:marginal_model}) include the mean, variance, range and nugget parameters $\{\mu, \sigma^2, \bm \gamma, \eta\}$ with $\bm \gamma=(\gamma_1,\gamma_2)$ being the range parameters in the kernel function in  Equation (\ref{equ:matern_5_2}). 
{As the number of pixels is large, we employed the maximum likelihood estimator (MLE) to estimate the parameters, as the result is similar to the robust estimation by marginal posterior mode \cite{Gu2018robustness}.}  Given the range and nugget parameters, the MLE of the mean and variance parameters has a closed-form expression: $\hat \mu=(\mathbf 1^T_N \mathbf {\tilde R}^{-1} \mathbf 1_N)^{-1} \mathbf 1^T_N \mathbf {\tilde R}^{-1} \mathbf y_v$, where 
$\hat \sigma^2=\frac{S^2}{N}$ with 
$S^2=(\mathbf y_v -\hat \mu \mathbf 1_N)^T\bm \tilde {\mathbf R}^{-1}(\mathbf y_v -\hat \mu \mathbf 1_N)$. The logarithm of the profile likelihood after plugging the MLE of the mean and variance parameters follows: 
\begin{equation}
 \log(L(\bm \gamma, \eta))= C -\frac{1}{2} \log(| \tilde {\mathbf R} | ) -\frac{N}{2} \log(S^2),
 \label{equ:log_lik}
\end{equation}
where $C$ is a constant not related to $(\bm \gamma, \eta)$. 

The range and nugget parameters are then estimated by maximizing the logarithm of the profile likelihood function:  
\begin{equation}
(\bm {\hat \gamma}, \hat \eta)=\argmax_{(\bm \gamma, \eta)} \log(L(\bm \gamma, \eta)). 
\end{equation}
Vectorize signal $\mathbf f_v =\mbox{Vec}(\mathbf F)$ from model (\ref{equ:model_matrix}). After plugging the MLE of the parameters, the posterior distribution of the image vector follows  
\begin{equation}
(\mathbf f_v \mid \mathbf y_v, \hat \mu, \hat \sigma^2 , \hat \eta, \hat{\bm \gamma}) \sim \mathcal{MN}( {\mathbf f}^*_v, \, \hat \sigma^2 \mathbf R^*  ), 
\label{equ:pred_dist}
\end{equation}
where the predictive mean and predictive covariance  follow
\begin{align}
 {\mathbf f}^*_v=&\hat \mu \mathbf 1_N+\mathbf R \tilde {\mathbf R}^{-1} (\mathbf y_v-\hat \mu \mathbf 1_N), \label{equ:pred_mean} \\
 \mathbf R^* =&  \mathbf R -  \mathbf R\tilde {\mathbf R}^{-1} \mathbf R. \label{equ:pred_cov}
\end{align}
 The predictive mean $ {\mathbf f}^*_v$ and predictive variances, the diagonal terms of $ \hat \sigma^2\mathbf R^*$, are required for prediction and uncertainty quantification.  
However, direct computation of the likelihood function involves inverting the covariance matrix, which takes $\mathcal O(N^3)$ operations, where the number of pixels $N$ can be at the order of $10^6-10^7$ for microscopy images. A fast computational way without  approximation is introduced in Section \ref{subsec:gp_images} to solve this computational challenge.

\subsection{Fast computation for  images}
\label{subsec:gp_images}
A wide range of approximation approaches of GPs have been developed \cite{snelson2006sparse,lindgren2011explicit,rue2009approximate,datta2016hierarchical,katzfuss2017multi,gramacy2015local,guinness2017circulant,kaufman2008covariance,zhu2024radial}. For GPs with product covariances on image data, no approximation is required. 
Denote the eigendecomposition of the subcovariance $\mathbf R_l= \mathbf U_l \bm \Lambda_l \mathbf U^T_l$, where $\mathbf U_l$ and $\bm \Lambda_l$ are matrices of eigenvectors and a diagonal matrix of eigenvalues for $l=1,2$, respectively. Furthermore, denote $\bm \lambda_l=(\lambda_{1,l},...,\lambda_{n_l,l})^T$, an $n_l$-vector of the diagonal terms in $\bm \Lambda_l $, 
for $l=1,2$. 

The log profile likelihood in Equation (\ref{equ:log_lik}) and the predictive distributions in Equation (\ref{equ:pred_dist}) can be written as a function in terms of eigenpairs in Lemma \ref{lemma:profile_likelihood} below. {{Lemma  \ref{lemma:profile_likelihood}  eliminates the need to invert the $N\times N$ covariance matrix and reduce the computational complexity to $\mathcal{O}(n_1^3+n_2^3)$, which is significantly smaller than $\mathcal{O}(n_1^3 n_2^3)$}}. The proof of Lemma \ref{lemma:profile_likelihood} is provided in the Appendix. 

\begin{lemma}\label{lemma:profile_likelihood}
\begin{enumerate}
\item (Profile likelihood). Denote the transformed likelihood vector $\tilde {\mathbf y}_v=\mbox{Vec}({\tilde {\mathbf Y} })= \mbox{Vec}(\mathbf U^T_1 (\mathbf Y -\hat \mu \mathbf 1_{n_1\times n_2}) \mathbf U_2)$. The logarithm of the profile likelihood in Equation (\ref{equ:log_lik}) can be written as  
\begin{align}
 \log(L(\bm \gamma, \eta))=& C -\frac{1}{2} \sum^{n_1}_{i=1} \sum^{n_2}_{j=1} \log\left(\lambda_{i,1} \lambda_{j,2}+ \eta \right)-\nonumber \\
 &\frac{N}{2} \log \left(\sum^{n_1}_{i=1} \sum^{n_2}_{j=1} \frac{\tilde  Y_{i,j}^2}{\lambda_{i,1} \lambda_{j,2}+ \eta} \right),
 \label{equ:log_lik_fast_eigen}
\end{align}
where 
\begin{align}
\hat \mu=&\left(\sum^{n_2}_{j=1} \sum^{n_1}_{i=1}\frac{\tilde u^2_{i,1} \tilde u^2_{j,2}}{\lambda_{i,1}\lambda_{j,2}+\eta} \right)^{-1}\times \nonumber 
\\
&\left( \sum^{n_2}_{j=1} \sum^{n_1}_{i=1}\frac{\tilde u_{i,1}\tilde Y_{i,j,0} \tilde u_{j,2}}{\lambda_{i,1}\lambda_{j,2}+\eta}\right),
\end{align}
with $\tilde Y_{i,j}$ being the $(i,j)$th entry in $(\mathbf U^T_1 (\mathbf Y -\hat \mu \mathbf 1_{n_1\times n_2}) \mathbf U_2)$, $\tilde Y_{i,j,0}$ being the $(i,j)$th entry of the $n_1\times n_2$ matrix $\mathbf U^T_1 \mathbf Y  \mathbf U_2$, ${\tilde u}_{i,1}$ being the $i$th term of the $n_1$ vector $\mathbf 1^T_{n_1} \mathbf U_1$ $i=1,...,n_1$ and ${\tilde u}_{j,2}$ being the $j$th term of the $n_2$ vector $\mathbf 1^T_{n_2} \mathbf U_2$, for $j=1,...,n_2$.  
\item (Predictive distribution).  
The predictive mean vector in Equation (\ref{equ:pred_mean}) follows 
\begin{align}
\mathbf f^*_v
=\hat \mu \mathbf 1_N+\mbox{Vec}( \mathbf U_1 \bm \Lambda_1  \mathbf{\tilde Y}_0 \bm \Lambda_2 \mathbf U^T_2),
\label{equ:fast_pred_mean}
\end{align}
where $\mathbf{\tilde Y}_0$ is a $n_1\times n_2$ with the $(i,j)$th entry being $\frac{\tilde{Y}_{i,j}}{\lambda_{i,1}\lambda_{j,2}+\eta}$ for $i=1,\dots, n_1$ and $j=1,\dots, n_2$. 
The predictive variance at the $(i,j)$th pixel follows
\[
\hat \sigma^2c^*_{i,j}=\hat \sigma^2 \left(1-\sum^{n_2}_{j'=1}\sum^{n_1}_{i'=1}\frac{\tilde r^2_{i,i',1} \tilde r^2_{j,j',2}}{\lambda_{i,1}\lambda_{j,2}+\eta}\right)\] where 
$\tilde r_{i,i',1}$ is the $i'$th term of the vector $\mathbf r^T_{i,1} \mathbf U_1$ and  $\tilde r_{j,j',2}$ is the $j'$th term of the vector $\mathbf r^T_{j,2} \mathbf U_2$, for $i'=1,...,n_1$ and   $j'=1,...,n_2$, with $\mathbf r^T_{i,1}=(K_1( x_{i,1}, x_{1,1}),...,K_1( x_{i,1}, x_{n_1,1}))^T$ and  $\mathbf r^T_{j,2}=(K_2( x_{j,2}, x_{1,2}),...,K_2( x_{j,2}, x_{n_2,2}))^T$. 
\end{enumerate}
\end{lemma}

\section{Image segmentation from Gaussian processes}
\label{sec:IS}

\begin{figure*}[t]
    \centering
        \includegraphics[width=1\linewidth]{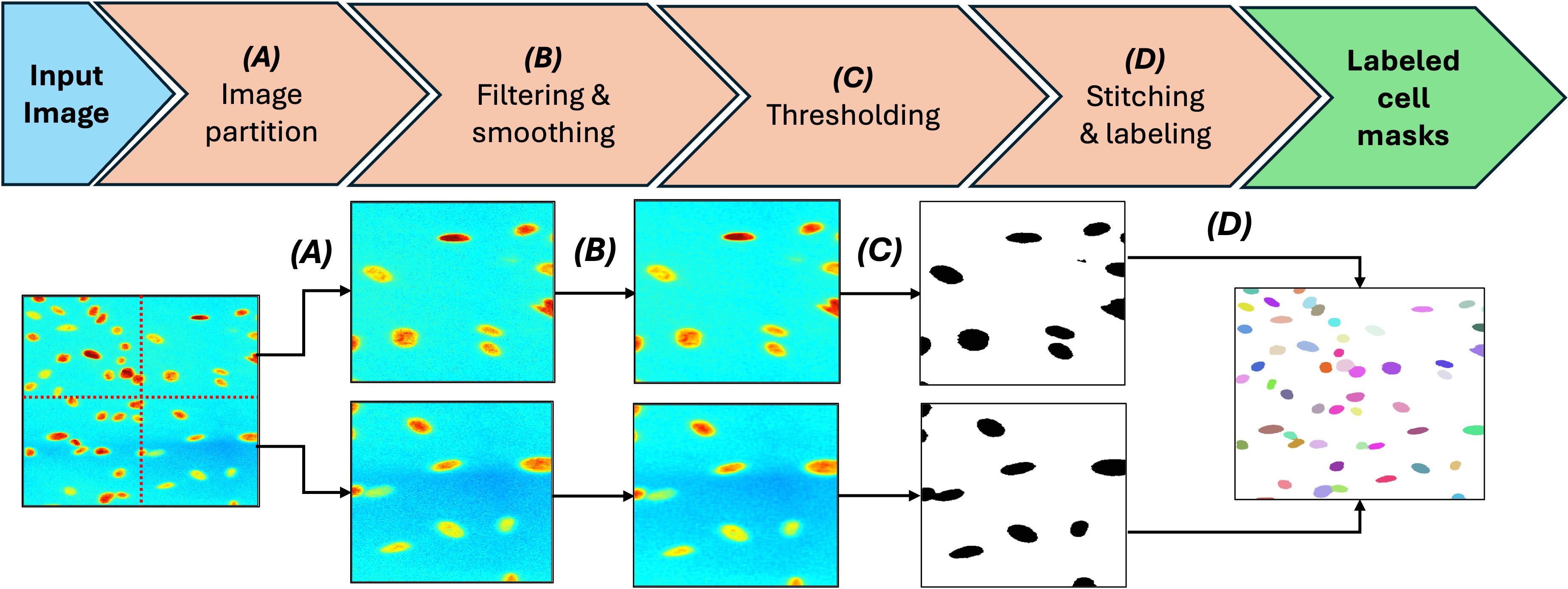}
    \caption{Workflow for segmenting and labeling cell images:
    {(A)} Divide the 
    image into different sub-images to enable locally estimated mean and variance parameters for capturing local properties such as the change of brightness. 
    {(B)} Compute the predictive mean of fast GPs in Section \ref{subsec:gp_images} to each sub-image, which     greatly reduces image noise. 
    {(C)} Threshold each smoothed sub-image based on the criterion discussed in Section \ref{subsec:crit_1}
    to produce binary images, separating cells from the background. The optimal threshold is estimated for each sub-image. {(D)} Recombine the binary sub-images into a single binary image, 
    and apply the watershed algorithm discussed in Section \ref{subsec:watershed_alg_sim} to the image for segmentation and labeling{, with each cell marked by a unique color.}
    }
    \label{fig:workflow}
\end{figure*}

Though Gaussian processes have been widely used in as surrogate models to emulate expensive computer simulations, nonparametric regression, and modeling spatially correlated data \cite{rasmussen2006gaussian,chang2016calibrating},the flexibility and scalability of GPs in unsupervised image segmentation have not been studied yet. 
The overall flow of our cellular image segmentation process is summarized in Figure \ref{fig:workflow}. 
First, some microscopy images of cells are particularly large with  $10^6-10^7$ pixels.
 Similar to other image segmentation methods \cite{schneider2012nih,stringer2021cellpose}, we segment each large image into multiple 
 sub-images. We then apply the fast algorithm of GPs in Lemma \ref{lemma:profile_likelihood} to get the predictive mean of each sub-image, illustrated in Parts (A) and (B) in Figure \ref{fig:workflow}, respectively. We assume the range parameters and nugget of the GPs are the same for each subimage estimated by MLE in Lemma \ref{lemma:profile_likelihood}, yet the mean and variance parameters are distinct for each separable image, and the estimation of these parameters has a closed-form expression. 
Such a step enables different estimated parameters to capture the change of optical properties, such as brightness difference and out-of-focus blur from distinct parts of a large microscopy image, which are crucial in practice.  
Second, we developed a robust measure that will be introduced in Section \ref{subsec:crit_1} to threshold the predictive mean of each sub-image into background and objects, as shown in Panel (C).  
Finally, the binary sub-images are restitched in Panel (D), and the watershed algorithm is applied to the binary matrix to retrieve the labeled cell masks, which will be introduced in Section \ref{subsec:watershed_alg_sim}.

\subsection{{ Image denoising} from Gaussian processes on lattice}
\label{subsec:image_denoising}

 Let $\mathbf{Y}$ represent the original cell image, which has size $N = n_1 \times n_2$.  We denote \(y(\mathbf{x}_{i,j})\) as the value of the image at pixel \(\mathbf{x}_{i,j} = (x_{i,1}, x_{j,2})\). {All possible pixel values are between 0 and 1, as each image in this analysis is in grayscale.} Microscopy images of cells can contain more than $10^6$ pixels, { and some areas of an image can have substantially higher pixel values than other areas}. Thus, for a large image $\mathbf{Y}$, it is normally partitioned into a set of cropped images, denoted as \(\{\mathbf{Y}_k\}_{k=1}^{\tilde K}\), where 
 $\tilde K$ is the total number of cropped images as illustrated in Figure \ref{fig:workflow}. { This step can substantially reduce the computational cost, and enable segmentation outcomes more adaptive to the features of the local areas in a large image. }

{ As experimental images often contain substantial noise}, for each cropped image, we first use GPs with fast computation introduced in Section \ref{subsec:lik_pred_GP} for denoising the images. The range and nugget parameters of GPs are estimated by maximizing the likelihood in Equation~(\ref{equ:log_lik_fast_eigen}), and they are
held the same for all sub-images as they have similar smoothness properties. The mean and variance parameters are estimated differently for each cropped image to capture the local change of the optical properties. As the sub-images will be turned into a binary image, the discontinuity of the boundary between different sub-images does not impact the image segmentation task.

\subsection{{ Generating} binary cell masks} 

\label{subsec:crit_1}

\begin{figure}[t]
 \centering    
 
    \includegraphics[width=.98\linewidth]{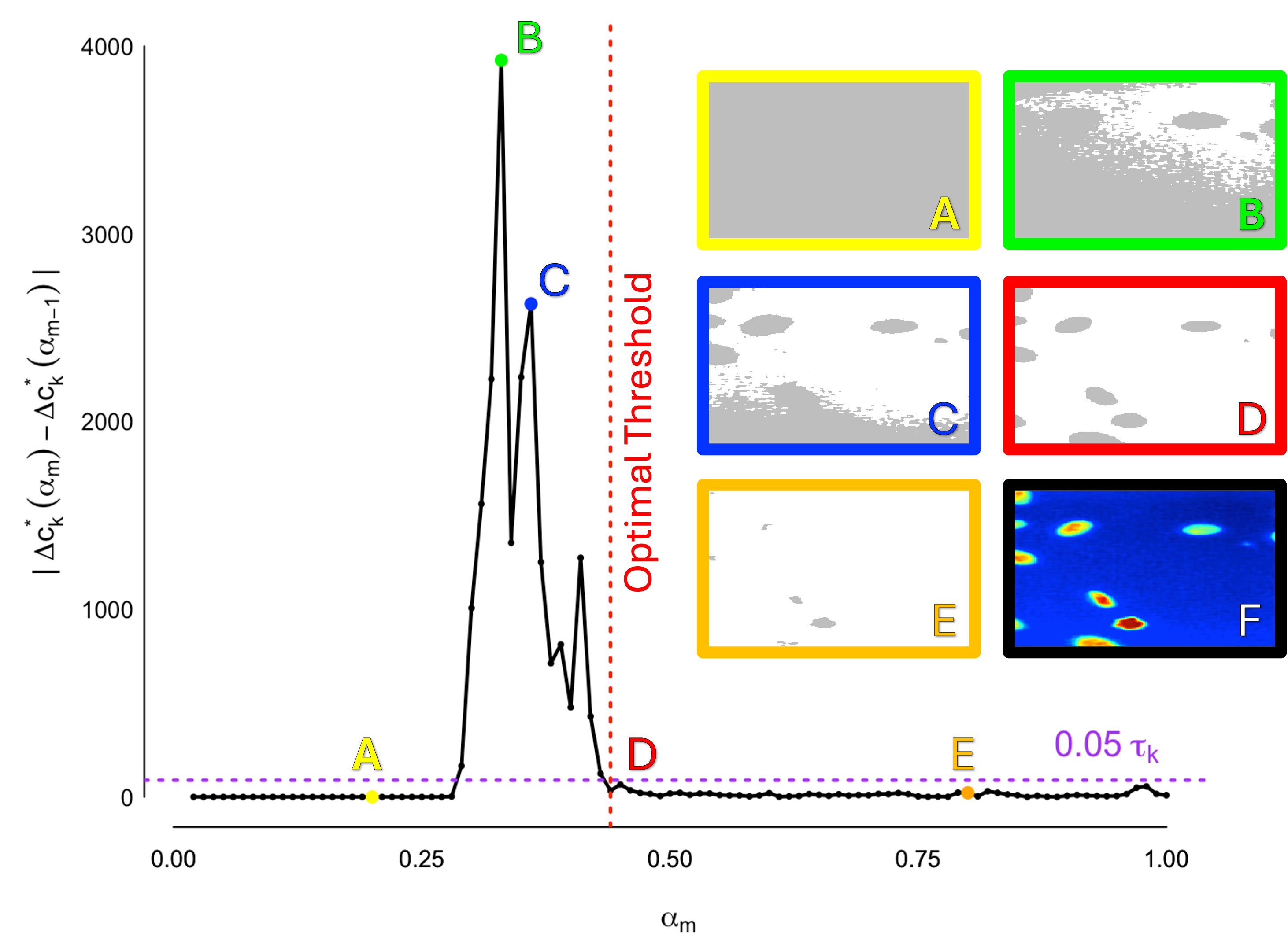}
    \caption{Comparison of binary image results across thresholds based on the absolute second difference in foreground pixels. The threshold, which ranges from 0-1, refers to the proportion of the maximum value of the predictive mean that is set as the binary cutoff. Images {A}, {B}, {C}, {D}, and {E} are the binary images generated by setting the corresponding threshold on cropped predictive mean image {F}. Note that the Image {B} corresponds with $\argmax_m \Delta c_{k}(\alpha_m)$ in Equation (\ref{eq:opt_threshold}).}
    \label{fig:crit_1_demonstration}
\end{figure}

The predictive mean at the interior pixel of a cell object typically has a larger value than the one in the background of a microscopy image. Inspired by the IsoData algorithm \cite{ridler1978picture} in {\sf ImageJ} \cite{abramoff2004image}, we developed an automated and robust way to determine the threshold value for separating objects from background noise using the predictive mean from GP models.  

Suppose each cropped image is $n_{k,1}\times n_{k,2}$ with predictive mean $\mathbf F^*_k$ with the $(i,j)$th entry being $F^*_k(\mathbf x_{i,j})$ and $\mbox{Vec}({\mathbf F^*_k})=\mathbf f^*_{k,v}$ given in Equation (\ref{equ:fast_pred_mean}). { As the intensity values of the pixels are between 0 to 1 and the predictive uncertainty is small, the range of $F^*_k(\mathbf x_{i,j})$ is also approximately between 0 to 1. 
 As the image has only one type of cells, the change of pixel values $F^*_k(\mathbf x_{i,j})$ is large if $\mathbf x_{i,j}$ is the pixel at the boundary of a cell object, and small elsewhere. }  
Thus we first compute the normalized predictive mean exceeds each threshold value: 
\[
c_{k}(\alpha_m) = \sum_{i,j} 
{1}_{\frac{{F}^*_{k}(\mathbf{x}_{i,j})}{\max(\mathbf{F}^*_{k})} > \alpha_m},
\] 
{ where  $\{\alpha_m\}^{M}_{m=1}$   a sequence of equally spaced thresholds from 0 to 1 with the default value $M=100$, and $\max(\mathbf{F}^*_{k})$ denotes the maximum value of the predictive mean $\mathbf{F}^*_{k}$ of the $k$th cropped image for $k=1,...,\tilde K$.  In general, the range of $\{\alpha_m\}^{M}_{m=1}$ can be chosen to be the range of $F^*_k(\mathbf x_{i,j})/\max(\mathbf{F}^*_{k})$.  } 
Subsequently, we calculate 
the difference \(\Delta c_k(\alpha_m)\) in pixel counts below
\[
\Delta c_{k}(\alpha_m) = {c_{k}(\alpha_{m-1}) - c_{k}(\alpha_{m})}.
\]
As $\Delta c_{k}(\alpha_m) $ may not be smooth over $\alpha_m$, we compute the predictive mean of $\Delta c_{k}(\alpha_m)$, denoted by $\Delta c^{*}_{k}(\alpha_m)$, via a GP model with the default Mat{\'e}rn kernel in Equation (\ref{equ:matern_5_2}) by the  $\sf{RobustGaSP}$ package \cite{gu2018robustgasp}.

 We define $\alpha^*$ as the smallest threshold $\alpha_m$ after the point of maximum fluctuation, denoted as $\arg \max_m \Delta c^*_{k}(\alpha_m)$, where the smoothed differences ${\Delta} c^*_{k}(\alpha_m)$ stabilize within a specified tolerance relative to the variability of $\Delta c_{k}(\alpha_m)$, an example of which is plotted in Figure \ref{fig:crit_1_demonstration}. Formally, let $\tau_k$ represent the standard deviation of the differences ${\Delta} c^*_{k}(\alpha_m)$. The optimal threshold $\alpha^*_{k}$ is given by:
\begin{align}
\label{eq:opt_threshold}
\alpha^*_k =& \min_m \{ \alpha_m : \left|{\Delta} c^*_{k}(\alpha_m) - {\Delta} c^*_{k}(\alpha_{m-1})\right| < 0.05\tau_k , \, \nonumber \\
&m > \arg \max_m \Delta c^*_{k}(\alpha_m) \}.
\end{align}

Figure \ref{fig:crit_1_demonstration} illustrates this process for determining the optimal threshold for segmenting cell objects from the background. While the first difference $\Delta c_k(\alpha_m)$  measures the immediate change in the number of pixels above each threshold, the second difference more precisely captures the stabilization process of these changes. Initially, at extremely small thresholds, almost all pixels are classified as foreground or cell objects, leading to minimal changes in pixel counts between threshold values, shown as A in Figure \ref{fig:crit_1_demonstration}. However, once the threshold starts to intersect the main background distribution, even small increments can trigger substantial variations, resulting in high second differences. As the threshold increases, the second difference reaches its maximum value at B in Figure \ref{fig:crit_1_demonstration} and then becomes erratic between B and C in Figure \ref{fig:crit_1_demonstration}. This erratic behavior occurs because, in this transitional range, as the predictive mean values near the decision boundary are very similar, even minor changes in the threshold result in disproportionately large shifts in the number of pixels classified as foreground or cell objects. Beyond this range, the second difference begins to stabilize and falls below a predetermined tolerance level show as E in Figure \ref{fig:crit_1_demonstration}, indicating that further increases yield only negligible changes in segmentation. Any further increase in the threshold 
results in over-segmentation. 

\begin{figure}[t]
 \centering
    \includegraphics[width=1\linewidth]{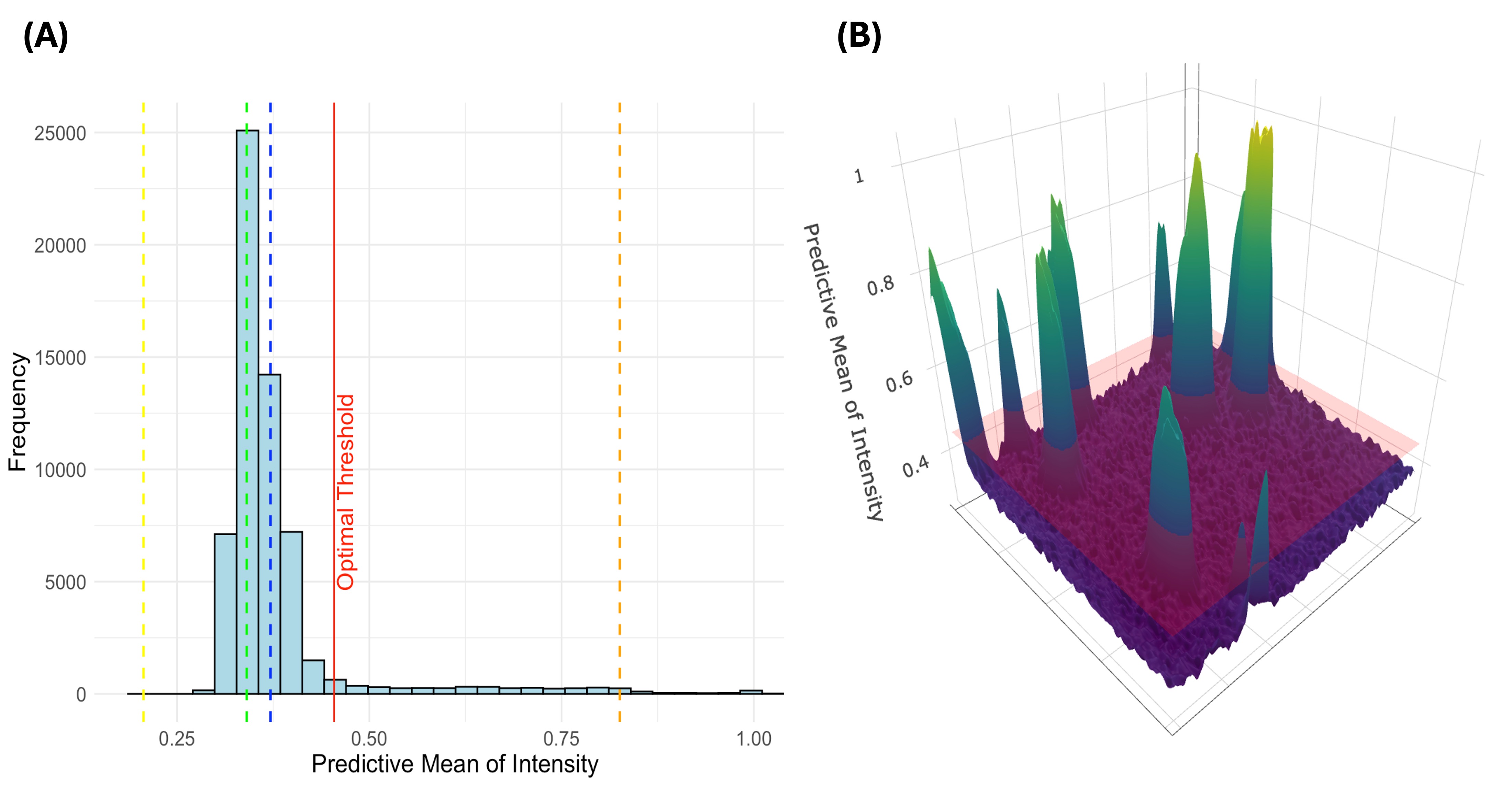}
    \caption{{(A)}  Frequency of the predictive mean of the intensity values over all pixels from the same microscopy image shown in Figure \ref{fig:crit_1_demonstration} {F}. The optimal threshold is annotated and lies right after the bulk of the background pixel values. The thresholds that generate images {A}, {B}, {C}, {D}, and {E} from Figure \ref{fig:crit_1_demonstration} are represented as vertical lines with the same color. All pixel intensity values less than the optimal threshold are background pixels and have a symmetric distribution. {(B)} The predictive mean is shown for each pixel with the optimal threshold plotted as the horizontal plane. 
    }
    \label{fig:crit_1_demo_histogram}
\end{figure}

Panel (A) in Figure \ref{fig:crit_1_demo_histogram} plots the location of different threshold values in the distribution of pixel intensity values. 
The optimal threshold selected by the algorithm lies on the right side of the mode. Panel (B) of Figure \ref{fig:crit_1_demo_histogram} plots a horizontal plane of the optimal threshold value. A threshold value much smaller than this value misclassifies the background as cell objects, whereas a threshold value much larger than the threshold value misses some small cell objects with low intensity peaks.

After estimating the threshold value \(\alpha^*_{k}\), we construct a binary image 
\(\mathbf{B}_{k}^*\) 
for each cropped image at each pixel: 
\[
{B}_{k}^*(\mathbf{x}_{i,j}) = 
\begin{cases}
1 & \text{if } {F}^*_{k}(\mathbf{x}_{i,j}) > \alpha^*_{k} \max(\mathbf{F}^*_{k}) \\
0 & \text{otherwise}
\end{cases}
\]
for  $k=1,...,\tilde K$. We complete panel (C) in  Figure \ref{fig:workflow} by generating binary cell masks for all sub-images. To address a rare scenario with a higher-than-expected object count, which only happened once for all sub-images, we implemented a re-thresholding step discussed in  Section S3
in the Supplementary Material. 
Finally, each object group is separated and detected via flood fill \cite{bhawnesh2020flood}, as described in  Section S4
in the Supplementary Material.

\subsection{{Labeling} cell masks by the watershed algorithm}
\label{subsec:watershed_alg_sim}
\begin{figure*}[t]
\centering
\includegraphics[width=1\textwidth]{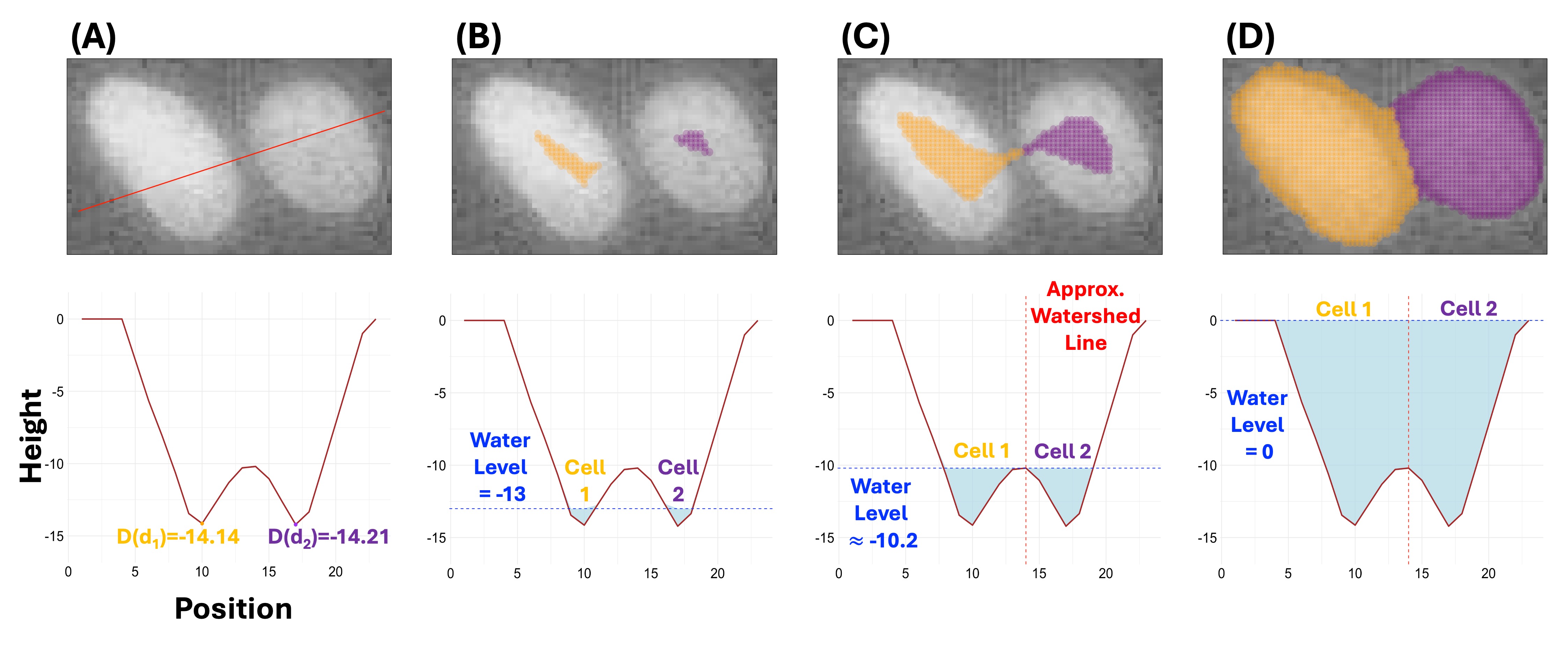}
\caption{An image of two cell nuclei (upper row) and heights of negative distances to the nearest background pixels (lower row). {(A)} Heights of negative distances along the red straight line in the upper panel are plotted in the lower panel before the watershed algorithm starts. 
{(B)} At water level $= -13$, both catch basins are partially filled, as both local minima are less than the current water level. The two separate water sources have unique labels and are not yet touching. {(C)} At water level at around $-10.2$, the water sources from the two catch basins flow into each other and the watershed line is formed at water level at around $-10.2$. ({D}) At water level $= 0$, the cell objects are filled with water, and the watershed operation is complete. Each cell is labeled and separated.
}
\label{fig:watershed_sim}
\end{figure*}

After obtaining the binary image, we apply the \texttt{watershed} algorithm to separate cell objects from \( \sf{EBImage} \) package in \( \sf{R} \) \cite{pau2010ebimage}, which is particularly useful for segmenting objects connecting to each other. 
The watershed algorithm creates a topological representation of an image based on a distance map, which is derived from the binary image representing foreground objects (cells).  
The distance map assigns a value to each foreground pixel based on its Euclidean distance to the nearest background pixel; higher values represent larger distances from the background. These distance values act as heights in a topological landscape, providing a basis for the watershed function to simulate water through. 
The negative distance map is then calculated by taking the negative of the distance map. Examples of the binary image and negative distance map are provided {in Section S5 
in the Supplementary Material.}

The local minima of the negative distance map 
generally correspond to the centers of cell objects and are used as the starting point for the ``flooding" process in the watershed algorithm.  The water fills the cell object from these local minima as the negative distance map is dunked into water, so the cell object acts as a catch basin as the map floods. As water from different starting points meets, watershed lines are formed, which delineate boundaries for distinct binary cell objects. Additionally, whenever the water from a catch basin meets a background pixel, a watershed line is formed \cite{Vincent1991}. These watershed lines separate and create distinct cell objects, and these cell objects can be assigned unique labels.

An example of the watershed algorithm at distinct water levels is provided in 
Figure \ref{fig:watershed_sim} with two cells denoted by $d_1$ and $d_2$.  
The top row of Figure \ref{fig:watershed_sim} plots the cell images where the negative heights of a slice over the red line are plotted in the bottom row. 
Figure \ref{fig:watershed_sim} (A) shows depths for the two local minima for the 1D example are $D(d_1)=-14.14$ and $D(d_2)=-14.21$. 
Each height acts as a water level, and the local minima act as seed points for water to flow through into each valley or catch basin. 
When the water level rises to touch a local minimum, the water is given the same label as the local minimum. As the water level rises, all unidentified pixels at or below the water level with neighboring labeled pixels are given the same label. 

Figure \ref{fig:watershed_sim} (B) shows the water levels at $-13$. 
Each catch basin is partially filled, as the water level is higher than the two local minima. 
As the water in each valley is still well separated, 
all emerged pixels in each valley are either labeled corresponding with $d_1$ or $d_2$, with no neighboring pixels containing a different label.

At water level around $-10.2$, shown in Figure \ref{fig:watershed_sim} (C),  the water from the two different basins begins to touch. 
The water from the two catch basins corresponding to $d_1$ and $d_2$ meet at the unlabeled pixel $p_0$ 
with a position close to 14, plotted in the x-coordinate.  The Euclidean distances between $p_0$ and the local minima are utilized to determine its assignment, and a watershed line is formed. The Euclidean distance between $p_0$ and $d_2$ ($\approx5.01$ units) is smaller than the distance between $p_0$ and $d_1$ ($\approx 5.62$ units), so $p_0$ is grouped with cell 2. The watershed line will serve as the boundary between the two cells. 

At the water level $0$, the objects are completely segmented, shown in Figure \ref{fig:watershed_sim} (D). 
The watershed operation then halts, as this height is used as a stopping point since all cell objects have negative heights in the negative distance map. 
Additional details of the watershed algorithm, including definitions and edge cases, can be found in \cite{Vincent1991}. 

Once all objects are segmented by watershed, the average foreground object pixel count is calculated. To filter out any noise that may have been included as a foreground object, all objects with pixel counts less than or equal to 15\% of the average are removed. If an object is touching the image boundary, the threshold count for removal is reduced to 5\% of the average, as some objects could be partially cut off during imaging.

\begin{figure*}[t]
    \centering
    \includegraphics[scale=0.7]{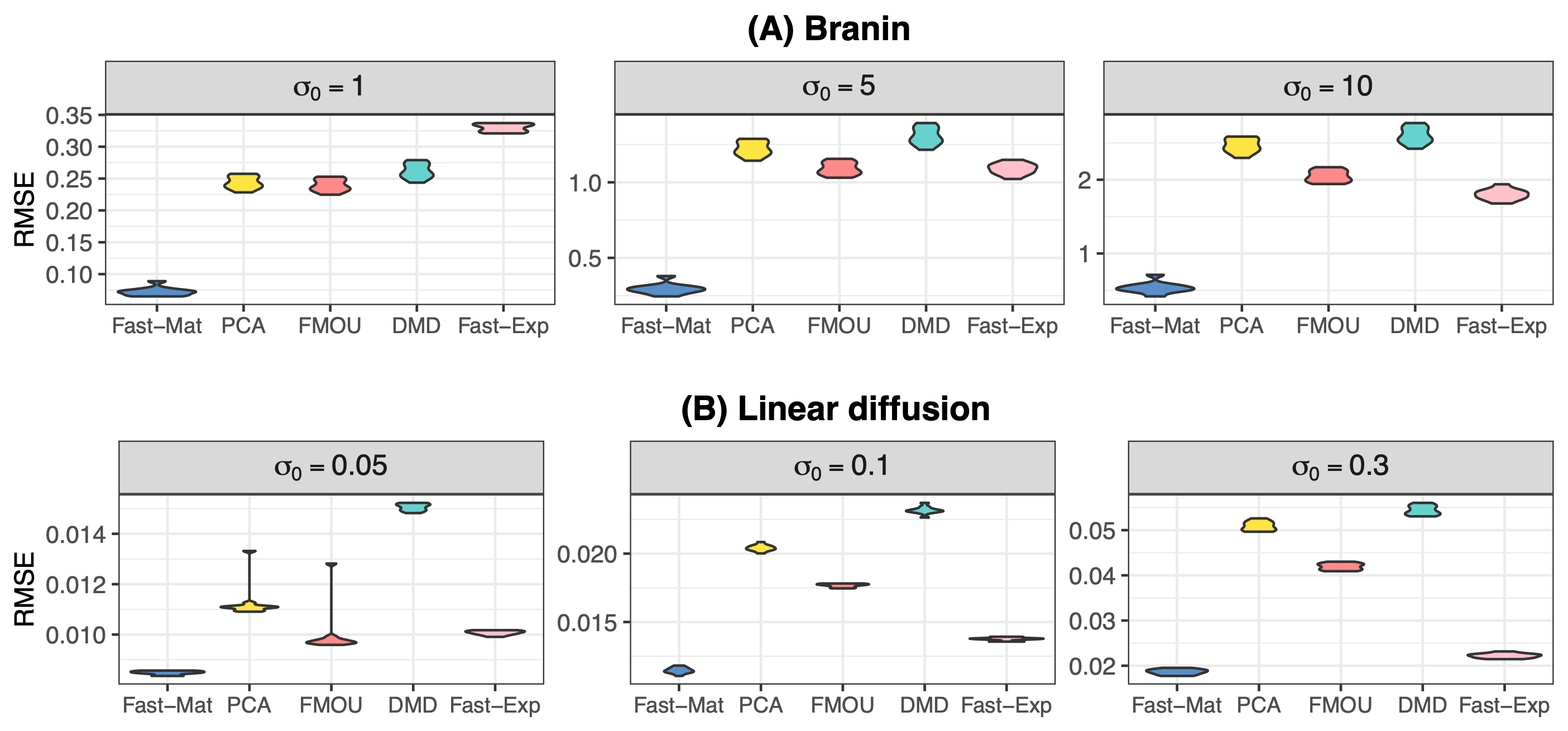}
    \caption{Violin plots of RMSE for five  methods applied to the Branin function and the linear diffusion equation across various noise levels. Each experiment  is repeated 10 times. The Fast-Mat and Fast-Exp represent the fast GPs with  Mat{\'e}rn kernels in Equation (\ref{equ:matern_5_2}) and exponential kernels, respectively.  
    }
    \label{fig:branin_linear_diffusion_rmse}
\end{figure*}

\section{Numerical studies of image denoising}
\label{sec:sim_study}

In this section, we numerically compare our fast GPs with several alternative methods by simulated experiments. The comparison includes the fast GP of a Mat{\'e}rn kernel with roughness parameter being 2.5 in Equation (\ref{equ:matern_5_2}) and the  
exponential kernels  $K_l(d_l)=\exp(-d_l/\gamma_l)$ where $d_l$ is  the distance of the $l$th coordinate of the input and  $\gamma_l$ is a range parameter, for $l=1,2$.
We also include three other alternative methods, the principal component analysis (PCA) \cite{berkooz1993proper}, fast algorithm of multivariate Ornstein-Uhlenbeck processes (FMOU) \cite{lin2025fast}, and dynamic mode decomposition (DMD) \cite{schmid2010dynamic}. PCA is commonly used in image denoising by projecting image data onto low-dimensional spaces spanned by the leading eigenvectors of the data covariance matrix. FMOU provides a fast expectation-maximization (E-M) algorithm for parameter estimation in a latent factor model $\mathbf Y = \mathbf U_0 \mathbf Z + \bm{\epsilon}$, where $\mathbf U_0$ is an orthogonal factor loading matrix and $\mathbf Z=[\mathbf z_1,\dots, \mathbf z_d]^T$ consists of $d$ independent latent processes, each modeled as an Ornstein-Uhlenbeck process with distinct correlation and variance parameters, and $\bm{\epsilon}$ are independent Gaussian noises. In each E-M iteration,  parameters are updated using closed-form expressions. 
DMD is popular in video processing to extract the spatiotemporal structure in nonlinear dynamical systems of high-dimensions. It assumes a linear transition matrix $\mathbf A$, estimated as $\hat{\mathbf A}=\argmin_\mathbf{A}||\mathbf Y_{2:n_2}-\mathbf{A}\mathbf{Y}_{1:n_2-1}||$, 
where $\mathbf Y_{2:n_2}$ and $\mathbf{Y}_{1:n_2-1}$ represent the last and first $n_2-1$ columns of $\mathbf Y$, respectively. 
An efficient algorithm, known as exact DMD \cite{tu2014dmd}, is proposed to identify the leading eigenpairs of $\mathbf{A}$ with computational complexity of $\min(\mathcal{O}(n_1 n_2^2),\mathcal{O}(n_1^2n_2))$. 

We study two examples. In the first example, we studied two scenarios, where the means of the observations follow the Branin function  \cite{picheny2013benchmark} and the linear diffusion equation \cite{lu2020prediction}, respectively. In the second example, the means of the observations are images of cell nuclei and whole cells. 

\begin{figure*}[t]
    \centering
    \includegraphics[scale=0.7]{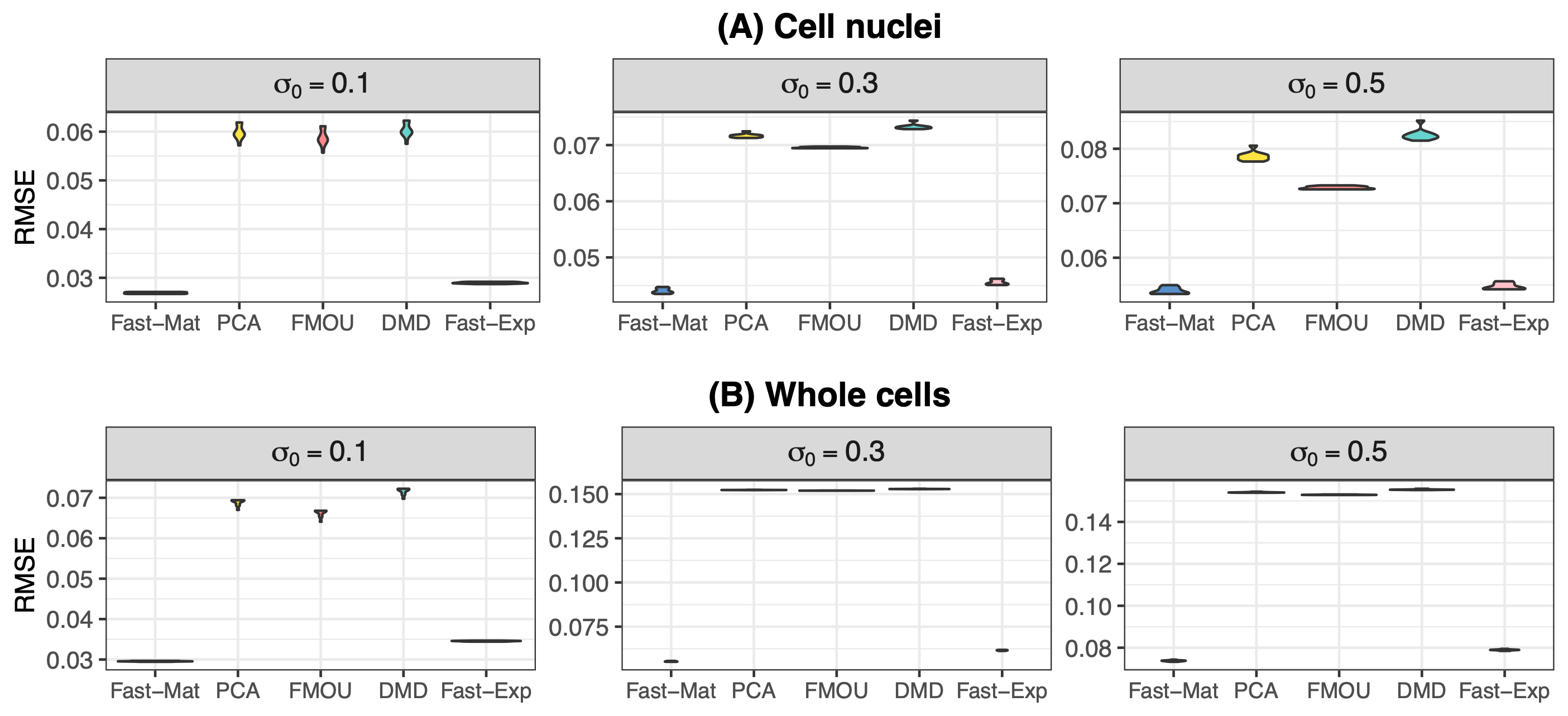}
    \caption{Violin plots of RMSE for five methods applied to images of noisy cell nuclei and whole cells across various noise levels. Each experiment is repeated 10 times. The Fast-Mat and Fast-Exp represent the fast GPs with  Mat{\'e}rn kernels in Equation (\ref{equ:matern_5_2}) and exponential kernels, respectively. }
    \label{fig:nuclei_whole_cell_rmse}
\end{figure*}

\begin{example}
\label{example:branin_linear_diffusion}
    In this example, we use the Branin function and linear diffusion equation to generate the mean of the images. 
    \begin{enumerate}
    \item (Branin function) The mean of the observation is defined by the Branin function
    \begin{align*}
    f(x_1,x_2) =& a(x_2-bx_1^2+cx_1-r)^2+ \\
    &  s(1-t)\cos(x_1)+s,
    \end{align*}
    where $x_1 \in [-5,10]$ and $x_2 \in [0,15]$. The default parameter values are: $a=1$, $b=\frac{5.1}{4\pi^2}$, $c=\frac{5}{\pi}$, $r=6$, $s=10$ and $t=\frac{1}{8\pi}$. The input domain is discretized into a uniform grid with $n_1=100$ points along the $x_1$-axis and $n_2=100$ points along the $x_2$-axis. The observations are generated by $y(x_1, x_2)=f(x_1, x_2)+\epsilon$, where $\epsilon$ is an independent Gaussian noise with the noise variance $\sigma_0^2 \in \{1^2, 5^2, 10^2\}$. 
    \item (Linear diffusion) The mean of the observation is governed by the partial differential equation $$\frac{\partial f(x,t)}{\partial t} = D\frac{\partial^2 f(x,t)}{\partial x^2},$$
    where $f(x,t)$ represents the concentration of the diffusing material at location $x$ and time $t$, and $D$ is the diffusion coefficient. We set $D=1$ and discretize the spatial domain $[0,1]$ into $200$ equally spaced grid points. The initial condition is set as $f(x,0)=0$, with a boundary condition at one end, maintaining a constant external concentration of 1. The signal is simulated over the time interval $t\in[0,0.2]$ using $200$ time steps, computed with a numerical solver \citep{soetaert2010solving}. The observations are generated by $y(x, t)=f(x, t)+\epsilon$, where $\epsilon$ is an independent Gaussian noise. We evaluate the model performance under three configurations with noise variances  $\sigma_0^2 \in \{0.05^2, 0.1^2, 0.3^2\}$.
\end{enumerate}
\end{example}

\begin{example}[Cell images]
\label{example:cell_images}
    We consider two noisy cell microscopy images \cite{luo2023molecular}: one for cell nuclei and the other for the whole cell. Independent Gaussian noise with  $\sigma_0^2 \in \{0.1^2, 0.3^2, 0.5^2\}$ is added to each image to generate the observations. 
\end{example}

To compare the model performance in noise filtering, we consider the error of the predictive mean of the noisy observations, quantified by the root mean squared error (RMSE):
$$
\text{RMSE} = \sqrt{\frac{\sum_{i=1}^{n_1}\sum_{j=1}^{n_2}(\hat{y}(\mathbf{x}_{i,j})-\E[y(\mathbf x_{i,j})])^2}{n_1n_2}},
$$
where $\hat{y}(\mathbf{x}_{i,j})$ is the estimated mean of $y(\mathbf x_{i,j})$. 

Figure \ref{fig:branin_linear_diffusion_rmse} presents the RMSE in estimating the mean, based on the data-generating processes described in Example \ref{example:branin_linear_diffusion}. Our fast GPs with the Mat\'{e}rn kernel in Equation (\ref{equ:matern_5_2}) consistently achieve the highest accuracy across varying noise levels in both scenarios. This is because GPs with separable kernels capture 
spatial dependencies in both input directions, which are crucial for modeling images. In contrast, PCA implicitly assumes that the prior in one direction is independent, as the corresponding probabilistic model assumes that the latent factors are distributed as a standard Gaussian distribution \cite{tipping1999probabilistic}. 
Though DMD and FMOU assume the latent factors are from Gaussian processes, which capture output correlation over one input dimension, the latent factor loading matrix is estimated without modeling the correlation over the other input by a kernel function. In comparison, Fast GPs with Mat\'{e}rn kernel directly model the correlation over two inputs through a product kernel, which induces better predictions of the signals. 
Additionally, since DMD implicitly assumes a noise-free model \cite{gu2024probabilistic}, 
leading to degraded performance as noise variance increases. Figure 
S6
in the Supplementary Material shows the predictive signal by Fast GPs with Mat\'{e}rn kernel from noisy observations generated by the Branin function and the linear diffusion equation, which demonstrates close alignment with the ground truth.

\begin{figure*}[t]
    \centering
    \includegraphics[scale=1.18]{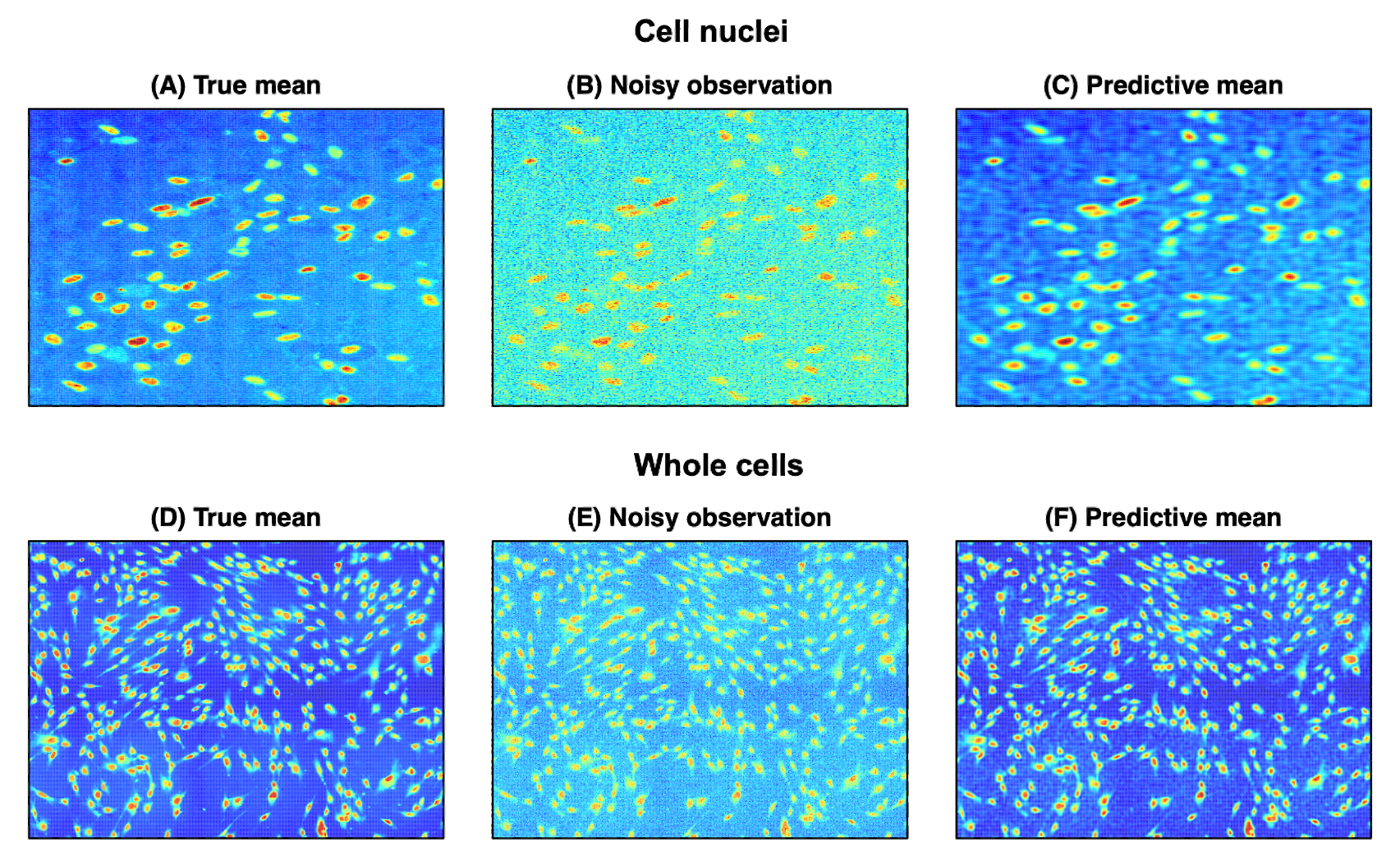}
    \caption{(A) True mean of cell nuclei. (B) Noisy observation of cell nuclei ($\sigma_0=0.1$). (C) Predictive mean of cell nuclei by fast GPs on lattice data with Mat\'{e}rn kernels in Equation (\ref{equ:matern_5_2}). (D) True mean of the whole cell. (E) Noisy observation of the whole cell ($\sigma_0=0.1$). (F) Predictive mean of the whole cell by Fast GPs on lattice data with  Mat\'{e}rn kernels. }
    \label{fig:nuclei_whole_cell_pred}
\end{figure*}

Figure \ref{fig:nuclei_whole_cell_rmse} compares the accuracy of five approaches in denoising cell nuclei and whole-cell images in Example \ref{example:cell_images}. The fast GPs with Mat\'{e}rn and exponential kernels significantly outperform PCA, FMOU and DMD across all settings, owing to their ability to estimate correlation parameters in both directions. Figure \ref{fig:nuclei_whole_cell_pred} plots the signal, noisy observations, and predictive mean from the fast GPs with a Mat\'{e}rn kernel in Equation (\ref{equ:matern_5_2}) for denoising the noisy cell images. The predictive mean is close to the latent mean of the observations. 

\begin{figure}[t]
    \centering
    \includegraphics[scale=0.7]{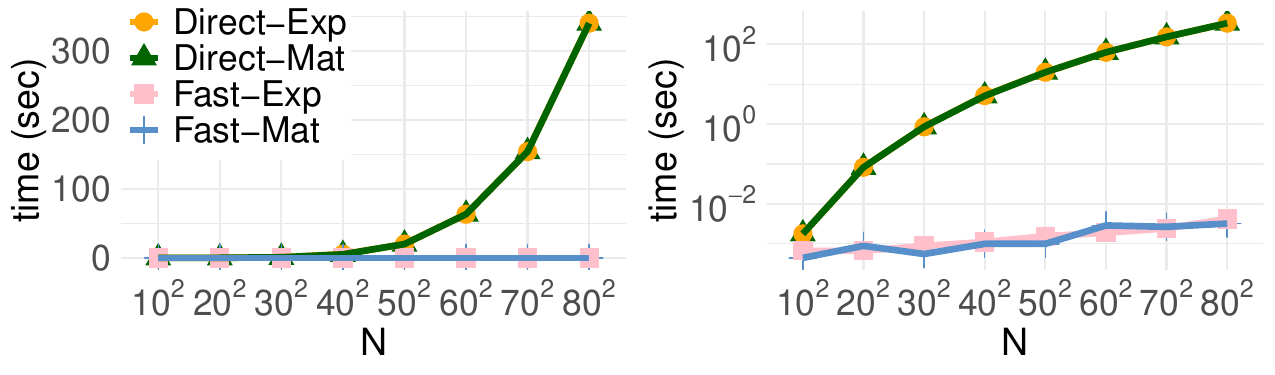}
    \caption{Comparison of the computational time in seconds for evaluating the profile likelihood of $\mathbf Y_v$ using direct and fast computation under varying sample sizes $N$. The  direct computation with the exponential kernel (Direct-Exp) is plotted as orange dots; the direct computation with the Mat\'{e}rn kernels in Equation (\ref{equ:matern_5_2}) (Direct-Mat) is plotted as green triangles;  3) the fast computation with the exponential kernel (Fast-Exp) is plotted as pink squares; The fast computation with the Mat\'{e}rn kernels in Equation (\ref{equ:matern_5_2}) (Fast-Mat) is plotted as blue crosses. The left panel shows the computational time in seconds (original scale), while the right panel displays the logarithmic scale.}
    \label{fig:compare_comp_time}
\end{figure}

Additionally, we compare the computational cost of evaluating the profile likelihood  by direct computation and fast computation. Simulations are conducted across varying sample sizes, ranging from $10^2$ to $80^2$. As shown in Figure \ref{fig:compare_comp_time}, the fast algorithm    in Equation (\ref{equ:log_lik_fast_eigen}) significantly reduces computational time compared to the direct approach, particularly for large sample sizes.

\section{Numerical studies of cell segmentation}
\label{sec:rda}

 \subsection{Criteria and methods}
We use microscopy images of human dermal fibroblasts (hDFs) 
in \cite{luo2023molecular} for testing different approaches. More details on the experimental conditions and techniques can be found in Section 
S7 
in the Supplementary Material. These images contain both nuclei and cytoplasm channels, and evaluation of the proposed GP-based segmentation method was performed on both channels.

 To evaluate the accuracy of the truth and segmentation results, we compare the area of overlap between the detected object and the true object relative to the overall union area of the two objects \cite{Rezatofighi_2019_CVPR}. Let $g$ represent a ground truth mask and $p$ a predicted mask. The Intersection over Union (IoU) metric between $g$ and $p$ is defined as:

\begin{equation}
\label{eq:iou}
    \text{IoU}(g, p) = \frac{|g \cap p|}{|g \cup p|}.
\end{equation}
A higher IoU indicates a better match between the predictive mask and true mask of an object, with the perfect match occurring at an IoU of 1. We test several levels of IoU for each method: 
$\alpha = [0.5, 0.55, 0.6, 0.65, 0.7, 0.75, 0.8]$. When $\alpha=0.6$, for instance, it means that whenever the IoU defined in Equation (\ref{eq:iou}) of a cell is larger than 0.6 by a certain method, this object will be classified as a true positive for this method. The range of thresholds allows us to assess image segmentation performance under distinct precision levels for defining true positives. 

 The average precision metric ($\text{AP}(\alpha)$) at each IoU threshold $\alpha \in [0,1]$ is often used for evaluating the correction of image segmentation \citep{stringer2021cellpose}.  For $\alpha\in [0,1]$,  
 a True Positive (TP) means that the ground truth mask $g$ is matched to a predicted mask $p$ with $\text{IoU}(g, p) \geq \alpha$. Otherwise, unmatched predicted masks are counted as False Positives (FP), and unmatched ground truth masks are counted as False Negatives (FN). The $\mbox{AP}(\alpha)$ is calculated as follows:

\begin{equation}
\label{eq:average_precison}
    \text{AP}(\alpha) = \frac{\text{TP}(\alpha)}{\text{TP}(\alpha) + \text{FP}(\alpha) + \text{FN}(\alpha)}. 
\end{equation}

 We compare {\sf ImageJ} segmentation and the GP-based unsupervised detection algorithm developed in this work. 
 We followed the conventional processing steps in {\sf ImageJ} to generate the cell masks  \cite{schneider2012nih}. All input images are converted to grayscale.
 To detect the foreground and background pixels, a threshold was set for each image using the default methods. To reduce noises in {\sf ImageJ}, a despeckle operation, which applies a $3 \times 3$ median filter, was then applied to each image. A watershed algorithm was used to ensure better separation of connecting cells. Finally, the cell masks were generated. 
 The cell size is selected to be larger than $30$ pixels. 
 Then, the binary image and the labeled mask can be generated in {\sf ImageJ}. 
 A detailed description of each step in this process for {\sf ImageJ} can be found in Section S8
 in the Supplementary Material.

\begin{figure*}[t]
    \centering
    \includegraphics[width=.98\textwidth]{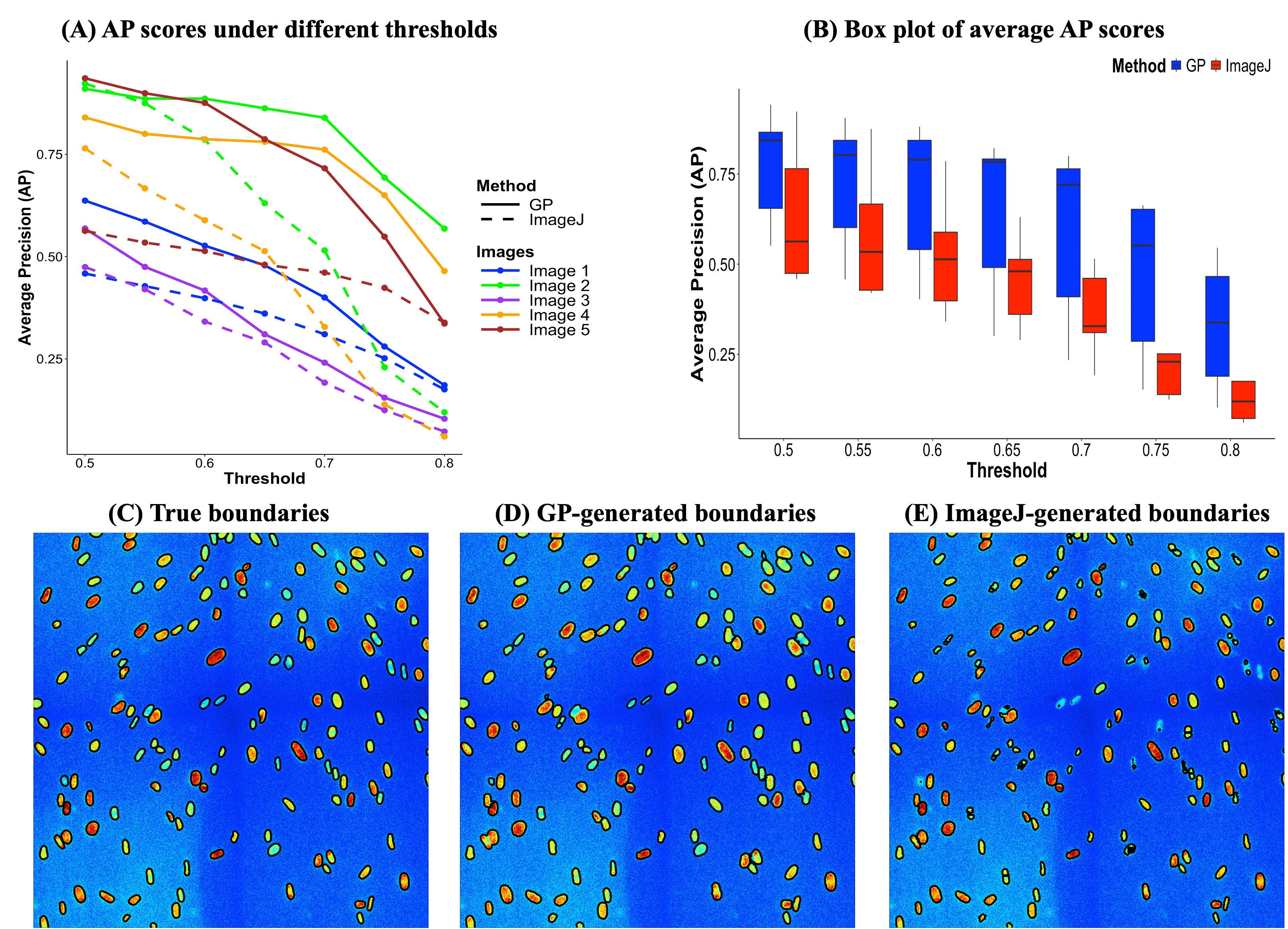} 
    \caption{
        Results for segmenting cell nuclei. {(A)} AP scores  for GP-based segmentation and {\sf ImageJ} across different thresholds. 
       {(B)} Boxplots of AP scores for each method. {These plots are created using the APs for each of the 5 images using each method across different thresholds.}
       {(C)} True boundaries of the fifth test image with $1024 \times 1024$ pixels. 
       {(D)} Boundaries generated by the GP-based method 
        for the fifth test image.
       {(E)} Boundaries generated using {\sf ImageJ} for the fifth test image. 
    }
    \label{fig:nuclei_cells_comparison}
\end{figure*}
\subsection{Image segmentation results for cell nuclei}

We first use 
five images of cell nuclei to test different methods. 
We choose a wide range of image sizes to test methods for images with different sizes. The smallest dimension of these images is $374\times 250$, and the largest image is $962\times1128$. The experimental details of these images are shown in Table S1 
in the Supplementary Material. 

In panel (A) of Figure~\ref{fig:nuclei_cells_comparison}, we plot the $\mbox{AP}(\alpha)$ between the GP-based segmentation and {\sf ImageJ} segmentation methods by solid lines and dashed lines, respectively, for 5 test images of nuclei at distinct threshold $\alpha$. 
For any image and threshold level, the solid line is almost always above the dashed line with the same color,  
as the GP-based method consistently achieves higher AP  than {\sf ImageJ} at any threshold level $\alpha$. The performance gap becomes more obvious at intermediate IoU thresholds between $0.6$ to $0.7$, which means that the GP-based method performs better at recovering the overall structure of the object than {\sf ImageJ}. 
The AP score of both methods drops at a high IoU threshold, as the annotated object may not be fully accurate, which can affect defining the truth. 
Panel (B) in Figure~\ref{fig:nuclei_cells_comparison}  shows the distribution of these AP scores from GPs and {\sf ImageJ} segmentation methods averaged at each threshold. Similar to the results in panel (A), the GP-based method achieves higher median AP scores across all thresholds. The difference between the GP-based method and  {\sf ImageJ} is more pronounced at a high threshold level, such as $\alpha=0.75$ or $\alpha=0.8$, where the boxes of two colors nearly have no overlap, indicating that GP-based method better capture the fine-scale boundary details than  {\sf ImageJ}. 

The true annotated boundaries, those generated by the GP-based method, and those generated by {\sf ImageJ} of the fifth test image are plotted in Figure~\ref{fig:nuclei_cells_comparison} (C)-(E). 
The GP-based image segmentation method generally produces smoother, more continuous boundaries capturing the actual cell nuclei, even in cases where cells were closely clustered. In contrast, the {\sf ImageJ} segmentation method sometimes produces fragmented boundaries and erroneously segmented individual nuclei into multiple smaller regions, especially in regions where the variations of pixel intensity are large. Additionally, {\sf ImageJ} fails to detect multiple cells entirely. Further image refinement may be achieved using additional plug-ins or filters from {\sf ImageJ}. However, this process would require extensive human intervention for trial and error, which would substantially increase the image processing time.

\subsection{Image segmentation results for whole cells}
\begin{figure*}[t]
    \centering
    \includegraphics[width=.98\textwidth]{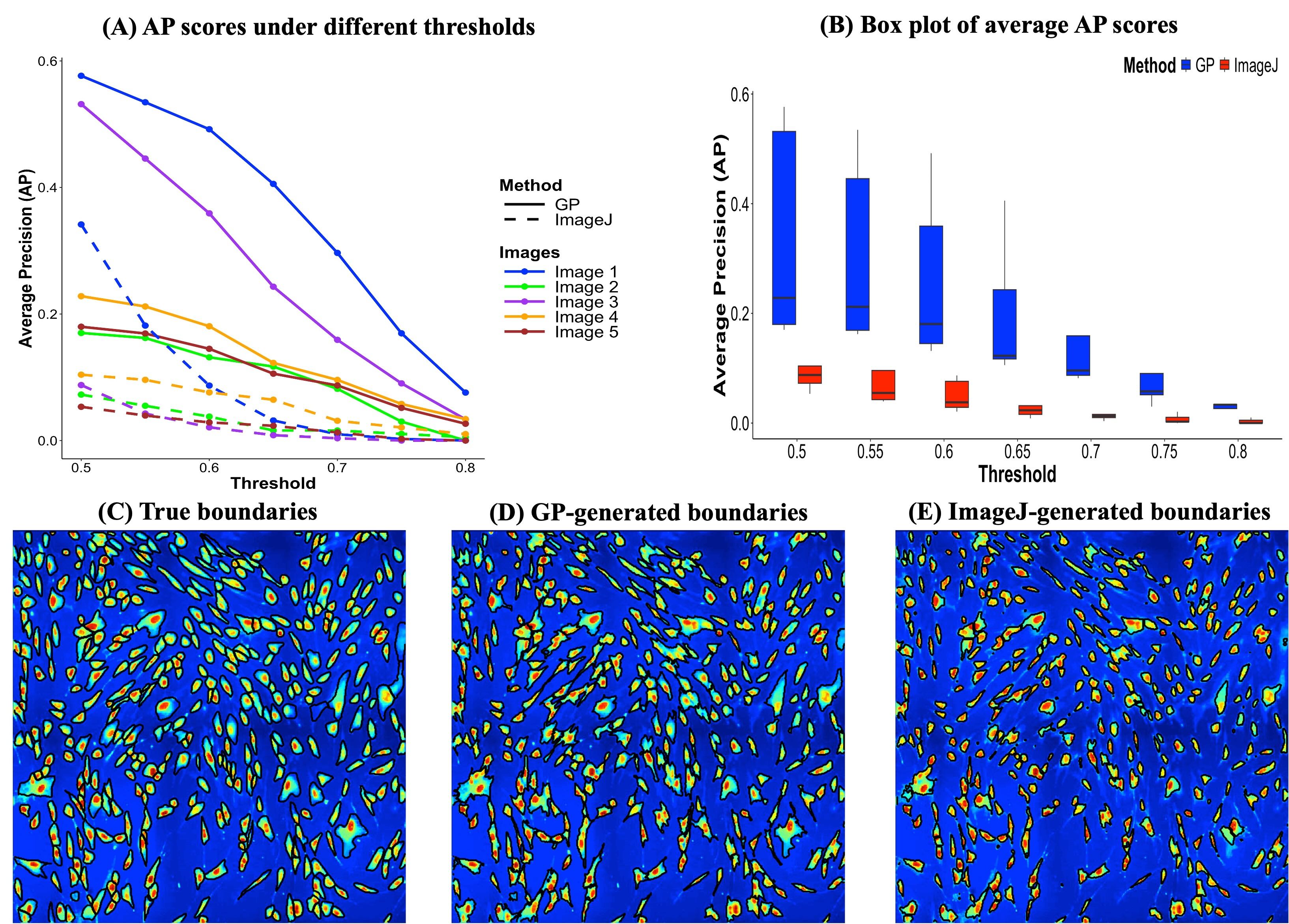}
    \caption{
        Results for segmenting whole cells. {(A)} AP scores for GP-based segmentation and {\sf ImageJ} across different thresholds. 
       {(B)} Boxplots of AP scores for each method.
       {(C)} True boundaries of the first test image with $602 \times 600$ pixels.
       {(D)} Boundaries generated by the GP-based method of the first test image. 
       {(E)} Same for panel (D) but generated using {\sf ImageJ}. 
    }
    \label{fig:whole_cells_comparison}
\end{figure*}

We compare the performance of the GP-based segmentation method and the {\sf ImageJ} method for five whole-cell images. 
The smallest image is $602\times 600$, and the largest is $1000\times1000$. 

Panel (A) of Figure~\ref{fig:whole_cells_comparison} shows the AP with different thresholds by the two methods for five images of whole cells. The variation of AP by both methods for the five images of whole cells is larger than the five images of the nuclei images. This is expected as segmenting whole cells is much more challenging due to the distinct shapes of the whole-cell images and more objects connecting to each other. Nonetheless, the solid line representing the AP for the GP-based method is again almost always above the dashed line representing the AP of the ImageJ segmentation results with the same color for all thresholds, meaning that the GP method consistently achieves higher AP scores than {\sf ImageJ}. 
Both methods show much lower precision at a smaller IoU threshold, as correctly identifying irregular boundary shapes in connecting whole cells is extremely hard to capture.  Such boundaries also pose challenges for humans to annotate, and for supervised methods to segment, even when annotated data are available. 

Panel (B) of Figure~\ref{fig:whole_cells_comparison} shows the distribution of the AP scores between the two methods. Compared to the segmentation results of the cell nuclei, the improvement by the GP-based segmentation methods is more pronounced for images of whole cells. The variation of AP by the GP-based segmentation is larger for whole-cell images than the ones for the images of nuclei, due to the difficulty in identifying the whole-cell shape either by statistical learning methods or by humans. 

The ground truth boundaries, the GP-based method generated boundaries, and the ImageJ generated boundaries of the first text image are given in panels (C)-(E) in Figure \ref{fig:whole_cells_comparison}. The GP-based method generally produces boundaries that better fit the actual cell shapes, preserving the irregular shapes of the cell cytoplasm. Although the GP-based method generally better preserves the true cell shapes, cell objects were also often grouped together. In contrast, {\sf ImageJ} segmentation often produces fragmented boundaries and splits individual cell shapes into multiple smaller regions.

\section{Concluding remarks}

In this study, we have presented a scalable cell image segmentation method for both the cell nucleus and cytoplasm. 
By leveraging fast Gaussian processes, automated thresholding and watershed operations, 
we have shown that the process is not only more automated, but the results are more precise 
than other unsupervised segmentation methods, such as {\sf ImageJ}. A  key strength of the proposed segmentation algorithm is that it does not rely on parameter tuning or labeled training datasets. The results of our unsupervised method demonstrate versatility across different cell channels, 
making it an appealing option for more general segmentation tasks beyond cell segmentation. 

Several future directions are worth exploring. First,  modern microscopes provide time-lapse videos for tracking the changes in cellular behavior over time. A future goal is to integrate our segmentation methods with particle tracking and linking algorithms that solve a linear assignment problem \cite{jaqaman2008robust}. 
As our approach does not restrict the shape of the object, it can be extended for image segmentation with objects undergoing dynamical changes, including microscopy images for the nucleation and growth process in crystallization \cite{wu2022nucleation} and protein dynamics \cite{lee2021myosin}. Second, when labeled image data are available, it is worthwhile to explore how these data can be used for constraining the shapes of the objects to improve estimation and simultaneously enable segmenting new objects not in the database. 
Third, the majority of cell segmentation methods are developed for 2D microscopy images. It is of interest to develop cell segmentation methods for microscopy that capture cell behaviors in a 3D environment, such as the orientational order and alignment \cite{huang2025cell}. { Similar to some popular cell segmentation tools, such as {\sf{Cellpose}} \cite{stringer2021cellpose}, our approach aims to segment images that contain the same type of cell. A future direction is to leverage techniques in tools such as {\sf{CellProfiler}} \cite{carpenter2006cellprofiler}  to segment images that contain multiple types of cells.} Furthermore, the uncertainty is often not quantified by the image segmentation methods, as the segmentation process often contains multiple steps, and the uncertainty from the convolutional neural network is an open question. As we model the image data with a probabilistic model, a future goal is to propagate the uncertainty throughout the analysis for uncertainty quantification.

\section*{Acknowledgement}
 This research is partially supported by the Materials Research Science and Engineering Center (MRSEC) by the National Science Foundation under Award No. DMR-2308708 and the Cyberinfrastructure for Sustained Scientific Innovation program by the National Science Foundation under Award No. OAC-2411043.  We thank  three anonymous referees for their comments that substantially improved this article.


\section*{Supplementary Material}
{The supplementary material provides additional details for image segmentation, experiments, and generation of ground truth.}

\begin{appendix}

\section*{Appendix: Proof of Lemma \ref{lemma:profile_likelihood}}
%


\begin{proof}
We first prove part 1 of the Lemma \ref{lemma:profile_likelihood}.  Given the eigendecompositions $
\mathbf R_l= \mathbf U_l \bm \Lambda_l \mathbf U^T_l$, where $\mathbf{U}_l \in \mathbb{R}^{n_l \times n_l}$ for $l=1,2$. Then the Kronecker product $\mathbf {\tilde R}= \mathbf R_2 \otimes \mathbf R_1+ \eta \mathbf I_{N}$ can be written as,
\begin{align*}
    \mathbf{\tilde R}  =& \mathbf U_2 \bm \Lambda_2 \mathbf U^T_2 \otimes \mathbf U_1 \bm \Lambda_1 \mathbf U^T_1 +\eta \mathbf I_{N}\\
    =& [(\mathbf U_2 \bm \Lambda_2) \otimes (\mathbf U_1\bm \Lambda_1)] (\mathbf U^T_2 \otimes \mathbf U^T_1) +\eta \mathbf I_{N}\\
    =& (\mathbf U_2 \otimes \mathbf U_1)(\bm \Lambda_2 \otimes \bm \Lambda_1) (\mathbf U^T_2 \otimes \mathbf U^T_1)+\eta \mathbf I_{N} \\
    =& \mathbf Q \bm {\tilde \Lambda}\mathbf Q^T,
\end{align*}
where $\mathbf Q = \mathbf U_2 \otimes \mathbf U_1$ is an orthogonal matrix, and $\bm {\tilde \Lambda}=(\bm \Lambda_2 \otimes \bm \Lambda_1+\eta \mathbf I_{N})$
is a diagonal matrix. 
Thus, 
\begin{align}
    |\tilde{\mathbf R}| = \prod^{n_1}_{i=1} \prod^{n_2}_{j=1} {(\lambda_{i,1} \lambda_{j,2}+ \eta)}.
    \label{equ:det}
\end{align}

Denote $\mathbf 1_N=\mathbf 1_{n_2} \otimes \mathbf 1_{n_1}$ and we have 
\begin{equation}
\hat \mu= \frac{\mathbf 1^T_N \mathbf {\tilde R}^{-1} \mathbf y_v} {\mathbf 1^T_N \mathbf {\tilde R}^{-1} \mathbf 1_N}.
\label{equ:mean_est}
\end{equation}
The numerator of Equation  (\ref{equ:mean_est}) can be written as 
\begin{align*}
    &\mathbf 1^T_N \mathbf {\tilde R}^{-1} \mathbf y_v  \\
    =& \mathbf 1^T_N(\mathbf Q \tilde{\bm \Lambda} \mathbf Q^T)^{-1}\mathbf y_v \\
    =& \mathbf 1^T_N(\mathbf U_2 \otimes \mathbf U_1) \tilde{\bm \Lambda}^{-1} \mathbf (\mathbf U_2 \otimes \mathbf U_1)^T \mbox{Vec}(\mathbf Y)\\
    =& [(\mathbf 1^T_{n_2} \mathbf U_2) \otimes (\mathbf 1^T_{n_1} \mathbf U_1)]\tilde{\bm \Lambda}^{-1} \mathbf (\mathbf U_2^T \otimes \mathbf U_1^T) \mbox{Vec}(\mathbf Y) \\
    =& ( \mathbf{ \tilde u}^T_2 \otimes \mathbf{ \tilde u}^T_1)\tilde{\bm \Lambda}^{-1} \mbox{Vec}(\mathbf U_1^T \mathbf Y \mathbf U_2) \\
    =& \sum^{n_2}_{j=1} \sum^{n_1}_{i=1}\frac{\tilde u_{i,1}\tilde Y_{i,j,0} \tilde u_{j,2}}{\lambda_{i,1}\lambda_{j,2}+\eta}.
\end{align*}
where $\tilde Y_{i,j,0}$ is the $(i,j)$th entry of the $n_1\times n_2$ matrix $\mathbf U^T_1 \mathbf Y  \mathbf U_2$,  ${\tilde u}_{i,1}$ is the $i$th term of $\mathbf {\tilde u}_1= \mathbf U^T_1 \mathbf 1_{n_1}$, and ${\tilde u}_{j,2}$ is the $j$th term of $\mathbf {\tilde u}_2= \mathbf U^T_2 \mathbf 1_{n_2}$, for $l=1,2$, $i=1,...,n_1$ and $j=1,...,n_2$. 

Similarly, the denominator of Equation (\ref{equ:mean_est}) can be written as 
\begin{align*}
    &\mathbf 1^T_N \mathbf {\tilde R}^{-1} \mathbf 1_N \\
    =& \mathbf 1^T_N(\mathbf U_2 \otimes \mathbf U_1) \tilde{\bm \Lambda}^{-1} \mathbf (\mathbf U_2^T \otimes \mathbf U_1^T)\mathbf 1_N \\
    =& [(\mathbf 1^T_{n_2} \mathbf U_2) \otimes (\mathbf 1^T_{n_1} \mathbf U_1)]\tilde{\bm \Lambda}^{-1} \mathbf [(\mathbf U_2^T \mathbf 1_{n_2}) \otimes (\mathbf U_1^T \mathbf 1_{n_1}) ] \\
    =& (\mathbf {\tilde u}^T_2 \otimes \mathbf {\tilde u}^T_1)\tilde{\bm \Lambda}^{-1} (\mathbf {\tilde u}_2 \otimes \mathbf {\tilde u}_1) \\
    =& \sum^{n_2}_{j=1} \sum^{n_1}_{i=1}\frac{\tilde u^2_{i,1} \tilde u^2_{j,2}}{\lambda_{i,1}\lambda_{j,2}+\eta}.
\end{align*}

The term $S^2$ in the log likelihood function in Equation (\ref{equ:log_lik}) follows
\begin{align}
    S^2 =& (\mathbf y_v -\hat \mu \mathbf 1_N)^T\bm \tilde {\mathbf R}^{-1}(\mathbf y_v -\hat \mu \mathbf 1_N) \nonumber \\
    =& (\mathbf y_v -\hat \mu \mathbf 1_N)^T \mathbf Q \tilde{\bm \Lambda}^{-1} \mathbf Q^T(\mathbf y_v -\hat \mu \mathbf 1_N)  \nonumber\\
    =& [\mathbf Q^T(\mathbf y_v -\hat \mu \mathbf 1_N)]^T \tilde{\bm \Lambda}^{-1} \mathbf Q^T(\mathbf y_v -\hat \mu \mathbf 1_N)  \nonumber\\
    =& [(\mathbf U_2^T \otimes \mathbf U_1^T)\mbox{Vec}(\mathbf Y -\hat \mu \mathbf 1_{n_1\times n_2})]^T  \nonumber \\
    &  \tilde{\bm \Lambda}^{-1}\times(\mathbf U_2^T \otimes \mathbf U_1^T)\mbox{Vec}(\mathbf Y -\hat \mu \mathbf 1_{n_1\times n_2}) \nonumber \\
    =& [\mbox{Vec}(\mathbf U_1^T(\mathbf Y -\hat \mu \mathbf 1_{n_1\times n_2}) \mathbf U_2)]^T  \\
    & \tilde{\bm \Lambda}^{-1}\nonumber\mbox{Vec}(\mathbf U_1^T(\mathbf Y -\hat \mu \mathbf 1_{n_1\times n_2}) \mathbf{U}_2)  \nonumber \\
    =& \tilde {\mathbf y}_v^T \tilde{\bm \Lambda}^{-1} \tilde {\mathbf y}_v \nonumber \\
    =& \sum^{n_1}_{i=1} \sum^{n_2}_{j=1} \frac{\tilde  Y_{i,j}^2}{\lambda_{i,1} \lambda_{j,2}+ \eta}, \label{equ:S_2}
\end{align}
where 
$\tilde {\mathbf y}_v=\mbox{Vec}(\mathbf U^T_1 (\mathbf Y -\hat \mu \mathbf 1_{n_1\times n_2}) \mathbf U_2)$. By Equations (\ref{equ:log_lik}), (\ref{equ:det}) and (\ref{equ:S_2}), we have  Equation (\ref{equ:log_lik_fast_eigen}). 

 For part 2 of the lemma,   we first  calculate the predictive mean vector below
    \begin{align*}
       \mathbf f^*_v =&\hat \mu \mathbf 1_N+ \mathbf R \mathbf {\tilde R}^{-1}(\mathbf y_v-\hat \mu \mathbf 1_N) \\
       =& \hat \mu \mathbf 1_N+(\mathbf U_2 \otimes \mathbf U_1)(\bm \Lambda_2 \otimes \bm \Lambda_1) (\mathbf U^T_2 \otimes \mathbf U^T_1) \times \nonumber \\
    &(\mathbf U_2 \otimes \mathbf U_1)\tilde{\bm \Lambda}^{-1} \mathbf (\mathbf U_2^T \otimes \mathbf U_1^T) (\mathbf y_v-\hat \mu \mathbf 1_N)\\
        =& \hat \mu \mathbf 1_N+(\mathbf U_2 \otimes \mathbf U_1)[(\bm \Lambda_2 \otimes \bm \Lambda_1) \tilde{\bm \Lambda}^{-1}] \times \nonumber \\
    &\mbox{Vec}(\mathbf U_1^T(\mathbf Y -\hat \mu \mathbf 1_{n_1\times n_2}) \mathbf{U}_2) \\
        =&\hat \mu \mathbf 1_N+(\mathbf U_2 \otimes \mathbf U_1)[(\bm \Lambda_2 \otimes \bm \Lambda_1) \tilde{\bm \Lambda}^{-1}] \tilde {\mathbf y}_v \\
         =& \hat \mu \mathbf 1_N+(\mathbf U_2 \otimes \mathbf U_1)\mbox{Vec}\left(\bm \Lambda_1 {\mathbf{\tilde Y}_0} \bm \Lambda_2 \right)\\
         =& \hat \mu \mathbf 1_N+\mbox{Vec}\left( \mathbf U_1\bm \Lambda_1 {\mathbf{\tilde Y}_0} \bm \Lambda_2 \mathbf U^T_2\right), 
    \end{align*}
    where $\mbox{Vec}(\mathbf{\tilde Y}_0)=\tilde{\mathbf \Lambda}^{-1}\tilde{\mathbf y}_v$.


To compute the predictive variance, the term $\mathbf R \mathbf {\tilde R}^{-1} \mathbf R$  follows

\begin{align*}
    \mathbf R \mathbf {\tilde R}^{-1} \mathbf R =&( \mathbf R_2 \otimes \mathbf R_1 )(\mathbf U_2 \otimes \mathbf U_1) \tilde{\bm \Lambda}^{-1} \mathbf (\mathbf U_2^T \otimes \mathbf U_1^T) \times \nonumber \\
    &(\mathbf R_2 \otimes \mathbf R_1 )\\
    =&( \mathbf R_2 \mathbf U_2\otimes \mathbf R_1 \mathbf U_1) \tilde{\bm \Lambda}^{-1} ( \mathbf U^T_2 \mathbf R_2 \otimes \mathbf U^T_1 \mathbf R_1 )
    \end{align*}
The $t=i+(j-1)n_1$ diagonal term of $\hat \sigma^2\mathbf R^*$ is the predictive variance at pixel location $(i,j)$, which follows  
\begin{align*}
\hat \sigma^2c^*_{i,j}=&\hat \sigma^2- \hat \sigma^2(\mathbf r^T_{j,2} \mathbf U_2 \otimes \mathbf r^T_{i,1} \mathbf U_1 ) \times \nonumber \\
&\tilde{\bm \Lambda}^{-1}  (\mathbf U^T_2 \mathbf r_{j,2} \otimes \mathbf U^T_1 \mathbf r_{i,1})  \\
=&\hat{\sigma}^2\left(1-\sum^{n_2}_{j'=1}\sum^{n_1}_{i'=1}\frac{\tilde r^2_{i,i',1} \tilde r^2_{j,j',2}}{\lambda_{i',1}\lambda_{j',2}+\eta}\right),
\end{align*}
where $\tilde r_{i,i',1}$ is the $i'$th term of the vector $\mathbf r^T_{i,1} \mathbf U_1$ and  $\tilde r_{j,j',2}$ is the $j'$th term of the vector $\mathbf r^T_{j,2} \mathbf U_2$, for $i'=1,...,n_1$ and   $j'=1,...,n_2$, with $\mathbf r^T_{i,1}=(K_1( x_{i,1}, x_{1,1}),...,K_1( x_{i,1}, x_{n_1,1}))^T$ and $\mathbf r^T_{j,2}=(K_2( x_{j,2}, x_{1,2}),...,K_2( x_{j,2}, x_{n_2,2}))^T$.

\end{proof}

\end{appendix}

  \bibliographystyle{abbrv} 
\bibliography{References_2024}

@article{pau2010ebimage,
  title={{EBI}mage—an {R} package for image processing with applications to cellular phenotypes},
  author={Pau, Gr{\'e}goire and Fuchs, Florian and Sklyar, Oleg and Boutros, Michael and Huber, Wolfgang},
  journal={Bioinformatics},
  volume={26},
  number={7},
  pages={979--981},
  year={2010},
  publisher={Oxford University Press}
}

@article{ridler1978picture,
  title={Picture thresholding using an iterative selection method},
  author={Ridler, Thomas Wilhelm and Calvard, S and others},
  journal={IEEE Trans. Syst. Man Cybern},
  volume={8},
  number={8},
  pages={630--632},
  year={1978}
}

@article{ferreira2011imagej,
  title={Image{J} user guide},
  author={Ferreira, Tiago and Rasband, Wayne},
  journal={USA: National Institutes of Health},
  year={2011}
}

@inproceedings{liu2009otsu,
  title={Otsu method and {K}-means},
  author={Liu, Dongju and Yu, Jian},
  booktitle={2009 Ninth International Conference on Hybrid Intelligent Systems},
  volume={1},
  pages={344--349},
  year={2009},
  organization={IEEE}
}

@article{luo2023molecular,
  title={Molecular-scale substrate anisotropy, crowding and division drive collective behaviours in cell monolayers},
  author={Luo, Yimin and Gu, Mengyang and Park, Minwook and Fang, Xinyi and Kwon, Younghoon and Urue{\~n}a, Juan Manuel and Read de Alaniz, Javier and Helgeson, Matthew E and Marchetti, Cristina M and Valentine, Megan T},
  journal={Journal of the Royal Society Interface},
  volume={20},
  number={204},
  pages={20230160},
  year={2023},
  publisher={The Royal Society}
}

@inproceedings{ronneberger2015u,
  title={U-net: Convolutional networks for biomedical image segmentation},
  author={Ronneberger, Olaf and Fischer, Philipp and Brox, Thomas},
  booktitle={Medical Image Computing and Computer-Assisted Intervention--MICCAI 2015: 18th international conference, Munich, Germany, October 5-9, 2015, proceedings, part III 18},
  pages={234--241},
  year={2015},
  organization={Springer}
}

@article{abramoff2004image,
  title={Image processing with {ImageJ}},
  author={Abr{\`a}moff, Michael D and Magalh{\~a}es, Paulo J and Ram, Sunanda J},
  journal={Biophotonics International},
  volume={11},
  number={7},
  pages={36--42},
  year={2004},
  publisher={Citeseer}
}

@article{tinevez2017trackmate,
  title={Track{M}ate: an open and extensible platform for single-particle tracking},
  author={Tinevez, Jean-Yves and Perry, Nick and Schindelin, Johannes and Hoopes, Genevieve M and Reynolds, Gregory D and Laplantine, Emmanuel and Bednarek, Sebastian Y and Shorte, Spencer L and Eliceiri, Kevin W},
  journal={Methods},
  volume={115},
  pages={80--90},
  year={2017},
  publisher={Elsevier}
}

@article{nichele2020quantitative,
  title={Quantitative evaluation of {ImageJ} thresholding algorithms for microbial cell counting},
  author={Nichele, Lorenzo and Persichetti, Valeria and Lucidi, Massimiliano and Cincotti, Gabriella},
  journal={OSA Continuum},
  volume={3},
  number={6},
  pages={1417--1427},
  year={2020},
  publisher={Optica Publishing Group}
}

@article{folkman1978role,
  title={Role of cell shape in growth control},
  author={Folkman, Judah and Moscona, Anne},
  journal={Nature},
  volume={273},
  number={5661},
  pages={345--349},
  year={1978},
  publisher={Nature Publishing Group UK London}
}

@article{gu2023data,
  title={Data-driven model construction for anisotropic dynamics of active matter},
  author={Gu, Mengyang and Fang, Xinyi and Luo, Yimin},
  journal={PRX Life},
  volume={1},
  number={1},
  pages={013009},
  year={2023},
  publisher={APS}
}

@article{eisenbarth2019dendritic,
  title={Dendritic cell subsets in T cell programming: location dictates function},
  author={Eisenbarth, SC},
  journal={Nature Reviews Immunology},
  volume={19},
  number={2},
  pages={89--103},
  year={2019},
  publisher={Nature Publishing Group UK London}
}

@article{meijering2012cell,
  title={Cell segmentation: 50 years down the road},
  author={Meijering, Erik},
  journal={IEEE Signal Processing Magazine},
  volume={29},
  number={5},
  pages={140--145},
  year={2012},
  publisher={IEEE}
}

@article{eisenhoffer2012crowding,
  title={Crowding induces live cell extrusion to maintain homeostatic cell numbers in epithelia},
  author={Eisenhoffer, George T and Loftus, Patrick D and Yoshigi, Masaaki and Otsuna, Hideo and Chien, Chi-Bin and Morcos, Paul A and Rosenblatt, Jody},
  journal={Nature},
  volume={484},
  number={7395},
  pages={546--549},
  year={2012},
  publisher={Nature Publishing Group UK London}
}

@article{liu2007cellular,
  title={Cellular and multicellular form and function},
  author={Liu, Wendy F and Chen, Christopher S},
  journal={Advanced Drug Delivery Reviews},
  volume={59},
  number={13},
  pages={1319--1328},
  year={2007},
  publisher={Elsevier}
}

@article{prasad2019cell,
  title={Cell form and function: interpreting and controlling the shape of adherent cells},
  author={Prasad, Ashok and Alizadeh, Elaheh},
  journal={Trends in Biotechnology},
  volume={37},
  number={4},
  pages={347--357},
  year={2019},
  publisher={Elsevier}
}

@article{bhawnesh2020flood,
author = {Bhawnesh, Kumar and Tiwari, Umesh and Kumar, Santosh and Tomer, Vikas and Kalra, Jasmeet},
year = {2020},
month = {08},
pages = {},
title = {Comparison and performance evaluation of boundary fill and flood fill algorithm},
volume = {8},
journal = {International Journal of Innovative Technology and Exploring Engineering},
doi = {10.35940/ijitee.L1002.10812S319}
}

@InProceedings{Rezatofighi_2019_CVPR,
author = {Rezatofighi, Hamid and Tsoi, Nathan and Gwak, JunYoung and Sadeghian, Amir and Reid, Ian and Savarese, Silvio},
title = {Generalized Intersection Over Union: A Metric and a Loss for Bounding Box Regression},
booktitle = {Proceedings of the IEEE/CVF Conference on Computer Vision and Pattern Recognition (CVPR)},
month = {June},
year = {2019}
}

@article{handcock1993bayesian,
  title={A {B}ayesian analysis of kriging},
  author={Handcock, Mark S and Stein, Michael L},
  journal={Technometrics},
  volume={35},
  number={4},
  pages={403--410},
  year={1993},
  publisher={Taylor \& Francis Group}
}

@article{tipping1999probabilistic,
  title={Probabilistic principal component analysis},
  author={Tipping, Michael E and Bishop, Christopher M},
  journal={Journal of the Royal Statistical Society: Series B (Statistical Methodology)},
  volume={61},
  number={3},
  pages={611--622},
  year={1999},
  publisher={Wiley Online Library}
}

@article{snelson2006sparse,
  title={Sparse {G}aussian processes using pseudo-inputs},
  author={Snelson, Edward and Ghahramani, Zoubin},
  journal={Advances in neural information processing systems},
  volume={18},
  pages={1257},
  year={2006},
  publisher={Citeseer}
}

@book{rasmussen2006gaussian,
  title={Gaussian processes for Machine Learning},
  author={Rasmussen, Carl Edward},
  year={2006},
  publisher={MIT Press}
}

@article{carpenter2006cellprofiler,
  title={Cell{P}rofiler: image analysis software for identifying and quantifying cell phenotypes},
  author={Carpenter, Anne E and Jones, Thouis R and Lamprecht, Michael R and Clarke, Colin and Kang, In Han and Friman, Ola and Guertin, David A and Chang, Joo Han and Lindquist, Robert A and Moffat, Jason and others},
  journal={Genome biology},
  volume={7},
  number={10},
  pages={1--11},
  year={2006},
  publisher={Springer}
}

@article{kaufman2008covariance,
  title={Covariance tapering for likelihood-based estimation in large spatial data sets},
  author={Kaufman, Cari G and Schervish, Mark J and Nychka, Douglas W},
  journal={Journal of the American Statistical Association},
  volume={103},
  number={484},
  pages={1545-1555},
  year={2008}
}

@article{rue2009approximate,
  title={Approximate {B}ayesian inference for latent {G}aussian models by using integrated nested {L}aplace approximations},
  author={Rue, H{\aa}vard and Martino, Sara and Chopin, Nicolas},
  journal={Journal of the royal statistical society: Series b (statistical methodology)},
  volume={71},
  number={2},
  pages={319--392},
  year={2009},
  publisher={Wiley Online Library}
}

@article{lindgren2011explicit,
  title={An explicit link between {G}aussian fields and {G}aussian Markov random fields: the stochastic partial differential equation approach},
  author={Lindgren, Finn and Rue, H{\aa}vard and Lindstr{\"o}m, Johan},
  journal={Journal of the Royal Statistical Society: Series B (Statistical Methodology)},
  volume={73},
  number={4},
  pages={423--498},
  year={2011},
  publisher={Wiley Online Library}
}

@article{schmid2010dynamic,
  title={Dynamic mode decomposition of numerical and experimental data},
  author={Schmid, Peter J},
  journal={Journal of Fluid Mechanics},
  volume={656},
  pages={5--28},
  year={2010},
  publisher={Cambridge University Press}
}

@article{roustant2012dicekriging,
   author = {Olivier Roustant and David Ginsbourger and Yves Deville},
   title = {DiceKriging, {DiceOptim}: Two {R} Packages for the Analysis of Computer Experiments by Kriging-Based Metamodeling and Optimization},
   journal = {Journal of Statistical Software},
   volume = {51},
   number = {1},
   year = {2012},
 issn = {1548-7660},   pages = {1--55},   doi = {10.18637/jss.v051.i01},
   url = {http://www.jstatsoft.org/index.php/jss/article/view/v051i01}
}

@article{schneider2012nih,
  title={{NIH} Image to {I}mage{J}: 25 years of image analysis},
  author={Schneider, Caroline A and Rasband, Wayne S and Eliceiri, Kevin W},
  journal={Nature methods},
  volume={9},
  number={7},
  pages={671--675},
  year={2012},
  publisher={Nature Publishing Group}
}

@article{picheny2013benchmark,
  title={A benchmark of kriging-based infill criteria for noisy optimization},
  author={Picheny, Victor and Wagner, Tobias and Ginsbourger, David},
  journal={Structural and Multidisciplinary Optimization},
  volume={48},
  number={3},
  pages={607--626},
  year={2013},
  publisher={Springer}
}

@article{tu2014dmd,
title = {On dynamic mode decomposition:  theory and applications},
journal = {Journal of Computational Dynamics},
volume = {1},
number = {2},
pages = {391-421},
year = {2014},
author = {Jonathan H. Tu and Clarence W. Rowley and Dirk M. Luchtenburg and Steven L. Brunton and J. Nathan Kutz},
}

@article{gramacy2015local,
  title={Local {G}aussian process approximation for large computer experiments},
  author={Gramacy, Robert B and Apley, Daniel W},
  journal={Journal of Computational and Graphical Statistics},
  volume={24},
  number={2},
  pages={561--578},
  year={2015},
  publisher={Taylor \& Francis}
}

@article{chang2016calibrating,
  title={Calibrating an ice sheet model using high-dimensional binary spatial data},
  author={Chang, Won and Haran, Murali and Applegate, Patrick and Pollard, David},
  journal={Journal of the American Statistical Association},
  volume={111},
  number={513},
  pages={57--72},
  year={2016},
  publisher={Taylor \& Francis}
}

@article{datta2016hierarchical,
  title={Hierarchical nearest-neighbor {G}aussian process models for large geostatistical datasets},
  author={Datta, Abhirup and Banerjee, Sudipto and Finley, Andrew O and Gelfand, Alan E},
  journal={Journal of the American Statistical Association},
  volume={111},
  number={514},
  pages={800--812},
  year={2016},
  publisher={Taylor \& Francis}
}

@article{guinness2017circulant,
  title={Circulant embedding of approximate covariances for inference from {G}aussian data on large lattices},
  author={Guinness, Joseph and Fuentes, Montserrat},
  journal={Journal of computational and Graphical Statistics},
  volume={26},
  number={1},
  pages={88--97},
  year={2017},
  publisher={Taylor \& Francis}
}

@article{katzfuss2017multi,
  title={A multi-resolution approximation for massive spatial datasets},
  author={Katzfuss, Matthias},
  journal={Journal of the American Statistical Association},
  volume={112},
  number={517},
  pages={201--214},
  year={2017},
  publisher={Taylor \& Francis}
}

@article{Gu2018robustness,
  title={Robust {G}aussian stochastic process emulation},
  author={Gu, Mengyang and Wang, Xiaojing and Berger, James O},
    journal={Annals of Statistics},
      volume={46},
  number={6A},
  pages={3038--3066},
    year    = {2018},

}

@article{gu2018robustgasp,
  author = {Mengyang Gu and Jesus Palomo and James O. Berger},
  title = {{RobustGaSP}: Robust {G}aussian Stochastic Process Emulation in
          {R}},
  year = {2019},
  journal = {{The R Journal}},
  doi = {10.32614/RJ-2019-011},
  pages = {112--136},
  volume = {11},
  number = {1}
}

@article{gu2022gaussian,
author = {Mengyang Gu and Hanmo Li},
title = {{G}aussian Orthogonal Latent Factor Processes for Large Incomplete Matrices of Correlated Data},
volume = {17},
journal = {Bayesian Analysis},
number = {4},
publisher = {International Society for Bayesian Analysis},
pages = {1219 -- 1244},
year = {2022},
doi = {10.1214/21-BA1295},
URL = {https://doi.org/10.1214/21-BA1295}
}

@article{stringer2021cellpose,
  title={Cellpose: a generalist algorithm for cellular segmentation},
  author={Stringer, Carsen and Wang, Tim and Michaelos, Michalis and Pachitariu, Marius},
  journal={Nature Methods},
  volume={18},
  number={1},
  pages={100--106},
  year={2021},
  publisher={Nature Publishing Group}
}

@article{lee2021myosin,
  title={Myosin-driven actin-microtubule networks exhibit self-organized contractile dynamics},
  author={Lee, Gloria and Leech, Gregor and Rust, Michael J and Das, Moumita and McGorty, Ryan J and Ross, Jennifer L and Robertson-Anderson, Rae M},
  journal={Sci. Adv.},
  volume={7},
  number={6},
  pages={eabe4334},
  year={2021},
  publisher={American Association for the Advancement of Science}
}

@article{gu2024probabilistic,
  title={Probabilistic forecast of nonlinear dynamical systems with uncertainty quantification},
  author={Gu, Mengyang and Lin, Yizi and Lee, Victor Chang and Qiu, Diana Y},
  journal={Physica D: Nonlinear Phenomena},
  volume={457},
  pages={133938},
  year={2024},
  publisher={Elsevier}
}

@article{lin2025fast,
  title={Fast data inversion for high-dimensional dynamical systems from noisy measurements},
  author={Lin, Yizi and Liu, Xubo and Segall, Paul and Gu, Mengyang},
  journal={arXiv preprint arXiv:2501.01324},
  year={2025}
}

@article{berkooz1993proper,
  title={The proper orthogonal decomposition in the analysis of turbulent flows},
  author={Berkooz, Gal and Holmes, Philip and Lumley, John L},
  journal={Annual review of Fluid Mechanics},
  volume={25},
  number={1},
  pages={539--575},
  year={1993},
  publisher={Annual Reviews 4139 El Camino Way, PO Box 10139, Palo Alto, CA 94303-0139, USA}
}

@article{lu2020prediction,
  title={Prediction accuracy of dynamic mode decomposition},
  author={Lu, Hannah and Tartakovsky, Daniel M},
  journal={SIAM Journal on Scientific Computing},
  volume={42},
  number={3},
  pages={A1639--A1662},
  year={2020},
  publisher={SIAM}
}

@article{soetaert2010solving,
  title={Solving differential equations in {R}: package de{S}olve},
  author={Soetaert, Karline ER and Petzoldt, Thomas and Setzer, R Woodrow},
  journal={Journal of Statistical Software},
  volume={33},
  number={9},
  year={2010},
  publisher={University of California Press}
}

@article{Vincent1991,
  author={Vincent, L. and Soille, P.},
  journal={IEEE Transactions on Pattern Analysis and Machine Intelligence}, 
  title={Watersheds in digital spaces: an efficient algorithm based on immersion simulations}, 
  year={1991},
  volume={13},
  number={6},
  pages={583-598},
  doi={10.1109/34.87344}}

@article{schindelin2012fiji,
  title={Fiji: an open-source platform for biological-image analysis},
  author={Schindelin, Johannes and Arganda-Carreras, Ignacio and Frise, Erwin and Kaynig, Verena and Longair, Mark and Pietzsch, Tobias and Preibisch, Stephan and Rueden, Curtis and Saalfeld, Stephan and Schmid, Benjamin and others},
  journal={Nature Methods},
  volume={9},
  number={7},
  pages={676--682},
  year={2012},
  publisher={Nature Publishing Group US New York}
}

@article{selmeczi2005cell,
  title={Cell motility as persistent random motion: theories from experiments},
  author={Selmeczi, David and Mosler, Stephan and Hagedorn, Peter H and Larsen, Niels B and Flyvbjerg, Henrik},
  journal={Biophysical journal},
  volume={89},
  number={2},
  pages={912--931},
  year={2005},
  publisher={Elsevier}
}

@article{tsukui2020collagen,
  title={Collagen-producing lung cell atlas identifies multiple subsets with distinct localization and relevance to fibrosis},
  author={Tsukui, Tatsuya and Sun, Kai-Hui and Wetter, Joseph B and Wilson-Kanamori, John R and Hazelwood, Lisa A and Henderson, Neil C and Adams, Taylor S and Schupp, Jonas C and Poli, Sergio D and Rosas, Ivan O and others},
  journal={Nature communications},
  volume={11},
  number={1},
  pages={1920},
  year={2020},
  publisher={Nature Publishing Group UK London}
}

@article{khang2025automated,
  title={Automated prediction of fibroblast phenotypes using mathematical descriptors of cellular features},
  author={Khang, Alex and Barmore, Abigail and Tseropoulos, Georgios and Bera, Kaustav and Batan, Dilara and Anseth, Kristi S},
  journal={Nature Communications},
  volume={16},
  number={1},
  pages={1--17},
  year={2025},
  publisher={Nature Publishing Group}
}

@article{barbazan2023cancer,
  title={Cancer-associated fibroblasts actively compress cancer cells and modulate mechanotransduction},
  author={Barbazan, Jorge and P{\'e}rez-Gonz{\'a}lez, Carlos and G{\'o}mez-Gonz{\'a}lez, Manuel and Dedenon, Mathieu and Richon, Sophie and Latorre, Ernest and Serra, Marco and Mariani, Pascale and Descroix, St{\'e}phanie and Sens, Pierre and others},
  journal={Nature Communications},
  volume={14},
  number={1},
  pages={6966},
  year={2023},
  publisher={Nature Publishing Group UK London}
}

@article{huang2025cell,
  title={Cell-Sheet Shape Transformation by Internally-Driven, Oriented Forces},
  author={Huang, Junrou and Chen, Juan and Luo, Yimin},
  journal={Advanced Materials},
  pages={2416624},
  year={2025},
  publisher={Wiley Online Library}
}

@article{wu2022nucleation,
  title={Nucleation and growth in solution synthesis of nanostructures--from fundamentals to advanced applications},
  author={Wu, Ke-Jun and Edmund, CM and Shang, Congxiao and Guo, Zhengxiao},
  journal={Progress in Materials Science},
  volume={123},
  pages={100821},
  year={2022},
  publisher={Elsevier}
}

@article{fang2024inverse,
  title={The inverse {K}alman filter},
  author={Fang, Xinyi and Gu, Mengyang},
  journal={arXiv preprint arXiv:2407.10089},
  year={2024}
}

@article{zhu2024radial,
  title={Radial neighbours for provably accurate scalable approximations of Gaussian processes},
  author={Zhu, Yichen and Peruzzi, Michele and Li, Cheng and Dunson, David B},
  journal={Biometrika},
  volume={111},
  number={4},
  pages={1151--1167},
  year={2024},
  publisher={Oxford University Press}
}

@article{jaqaman2008robust,
  title={Robust single-particle tracking in live-cell time-lapse sequences},
  author={Jaqaman, Khuloud and Loerke, Dinah and Mettlen, Marcel and Kuwata, Hirotaka and Grinstein, Sergio and Schmid, Sandra L and Danuser, Gaudenz},
  journal={Nature methods},
  volume={5},
  number={8},
  pages={695--702},
  year={2008},
  publisher={Nature Publishing Group US New York}
}


\end{document}